\documentclass{article}

\usepackage{authblk}
\usepackage[utf8]{inputenc}
\usepackage[left=2cm,right=2cm,top=1.5cm,bottom=1.5cm]{geometry}
\usepackage{amsmath}
\usepackage{amssymb}
\usepackage{graphicx}
\usepackage{subfig}
\usepackage{mathtools}
\usepackage{mathpazo}
\usepackage[mathpazo]{flexisym}
\usepackage{breqn}
\usepackage{verbatim}
\usepackage{xcolor}
\usepackage{placeins}
\usepackage{hyperref}
\usepackage{soul}
\usepackage{ulem}
\hypersetup{
    colorlinks=true,
    linkcolor=blue,
    filecolor=magenta,      
    urlcolor=cyan,
    citecolor=teal
    }
\usepackage{comment}
\graphicspath{{figs/}}

\usepackage[style=numeric-comp,style=phys,url=true]{biblatex}
\DeclareMathOperator{\sech}{sech}

\title{Kink movement on a periodic background}
\author[1]{Tomasz Dobrowolski}
\author[2]{Jacek Gatlik}
\author[3]{Panayotis G. Kevrekidis}
\affil[1]{\textit{Department of Computer Physics and Quantum Computing, University of the National Education Commission in Krakow, Podchor\c{a}\.zych  2, 30-084 Cracow, Poland}}
\affil[2]{\textit{AGH University of Krakow, Faculty of Physics and Applied Computer Science, 30-059 Krakow, Poland}}
\affil[3]{\textit{Department of Mathematics and Statistics, University of Massachusetts, Amherst, Massachusetts 01003-4515, USA}}
\date{\today}

\begin{document}
\maketitle

\begin{abstract}
The behavior of the kink in the sine-Gordon (sG) model in the presence of periodic inhomogeneity is studied. An ansatz is proposed that allows for the construction of a reliable effective model with two degrees of freedom. Effective models featuring very good agreement with the original field-theoretic partial differential equation are constructed, including in the non-perturbative region and for relativistic velocities.  
The numerical solutions of the sG model describing the evolution of the kink in the presence of a barrier as well as in the case of a periodic heterogeneity under the potential additional influence of a switched bias current and/or dissipation were obtained. The results of the field equation and the effective models were favorably compared.
The effect of the choice of initial conditions in the field model on the agreement of the results with the effective model is also discussed.
\end{abstract} \hspace{10pt}

\section{Introduction}
Solitons, which are solutions of some nonlinear field equations, were introduced to mathematical physics in the context of Korteweg de Vries equation \cite{Zabusky1965}.
motivated from diverse physical applications
including, e.g.,
the behaviour of solitary waves on shallow water channels, as well as the continuum limit of
prototypical lattice models such as the 
Fermi-Pasta-Ulam-Tsingou model~\cite{FPU55,Zabusky1965}.  The existence of solitons is possible due to the balance between nonlinearity and dispersion.  It soon became apparent that other equations had similar solutions. Among such dispersive nonlinear
partial differential equation (PDE) models, the sine-Gordon model plays a significant role~\cite{CKW14}. Remarkably, it has its own connections to continuum limits of
discrete systems in the form damped-driven
coupled torsion pendulum arrays~\cite{Cuevas2009}.
Yet, it also provides a basis for describing many continuum physical systems. One example is liquid crystals \cite{Lam1992}. In these materials, the field variable describes the orientation of the molecules. Other examples are quasi-one dimensional ferromagnetics or ferroelectrics \cite{Zharnitsky1998, Bugaychuk2002}. 
Arguably, however, the most well-studied
physical system described by the sine-Gordon model 
involves Josephson junctions. The latter is a relatively simple device consisting of two superconductors separated by a very thin insulator layer~\cite{Malomed2014}. This device is known for its many practical applications \cite{Braginski2019}.  To achieve the desired properties of the junction, various modifications of its shape 
have been proposed \cite{Monaco2016, Benabdallah1996, Kemp2002b, Gulevich2006}.

Our aim herein is to study a perturbed sine-Gordon model containing a deformation of the term describing the spatial variation of the {dispersion coefficient, under the
potential presence of external
drive, as well as loss in the
system.} 
Perturbations of this type have appeared in the context of curved Josephson junctions~\cite{Dobrowolski2012, Gatlik2021}. {Interestingly,
this model has also been proposed in connection to applications 
in the context of  the three and five-layer ferromagnetic structures \cite{Ekomasov2015, Ekomasov2016, Gumerov2019}}
The influence of deformations belonging to this type on the evolution of the kink was studied, for example, in the work \cite{Gatlik2023}. Effectively, this plays
the role of an effective potential barrier.  In particular, 
the presence of a forcing in the system in the form of bias current and also dissipation leads to an intriguing interplay related to the appearance of stable points in the model (depending on the value of dissipation and current) in the phase space of the system.
{In general, there are many studies devoted to the effect of junction geometry on fluxon dynamics. These works describe curved large area Josephson junctions \cite{Gulevich2007, Gorria2004}. Theoretically \cite{Goldobin2001} and experimentally \cite{Carapella2002}, the Josephson vortex ratchet design
has been considered. Moreover, in the context of vortex qubit preparation, the heart-shaped junctions were studied in articles \cite{Kemp2002a, Kemp2002b}. Nontrivial geometry is also crucial in the construction of an elementary flux quanta generator \cite{Gulevich2008}.}

The perturbed sine-Gordon model was also studied in the presence of inhomogeneities that were spatially periodic in nature \cite{Sanchez1992,Scharf1992}. Some of these inhomogeneities were in the character of microshorts and microresistors and are described in the classic work of~\cite{McLaughlin1978}. Research on this system was also conducted in the work of \cite{Malomed1988}, where the minimum bias current density that causes a pinned kink to move was determined, among other things. A slightly different analytical form of the periodic part of the potential was adopted, for example, in the work \cite{Mkrtchyan1979}. Moreover, an experimental study of the Josephson junction described by this model was carried out and presented in \cite{Serpuchenko1987}.  Other periodic perturbations of the sine-Gordon model described in the literature were in the nature of regular lattices of dissipative inhomogeneities \cite{Kivshar1988}. This work determined, among other things, the average kink velocity.

In the current work, we focus on studying the impact of periodic inhomogeneities, which could be interpreted in the context of a Josephson junction as periodic deformations of its shape. The work consists of the following parts. In the second section, we present the field model under study. The description also includes a presentation of the non-standard initial conditions used during the simulations carried out in the context of the field model. The third section is devoted to the construction of the effective description in the conservative system i.e. when there is neither dissipation nor bias current in the system.
This section includes a comparison of the effective model with the full field-theoretic PDE in the case of interaction of a kink with a barrier-like admixture and with periodic inhomogeneities.  The fourth part is devoted to a discussion of the choice of initial conditions in the field model. 
The fifth section contains the derivation of the effective equations of motion on the ground of the 
so-called non-conservative Lagrangian method~\cite{Galley2013,Kevrekidis2014}. This approach allows for the inclusion of dissipation and bias current in the description of the system. This section also contains examples to compare the results of the effective model with the original field model. The 
final section summarizes our findings and
presents our conclusions, as well as a number of
possible directions for future study.

\section{Model Setup}
We study the perturbed sine-Gordon (sG) model defined by the equation
\begin{equation}
\label{sine-gordon-dis}
\partial_t^2 \phi + \alpha \partial_t \phi - \partial_x (\mathcal{F}(x)\partial_x \phi) + \sin \phi = -\Gamma,
\end{equation}
where
\begin{equation}
\label{inhomogeneity}
\mathcal{F}(x) = 1 +\varepsilon g(x)
\end{equation}
describes the spatial heterogeneity of the system.
Here $\alpha$ stands for the disipation constant, while $\Gamma$ can be understood as a bias current. In this work, we will assume that this current can vary with time. The function $g$ appearing in the equation characterizes the departure from the homogeneous
spatial setting. In most of the study, we assume that this function is periodic in space. An additional important component of exploring
the system dynamics concerns the selection of appropriate initial conditions. In the present work, these conditions take the form of a moving kink (inspired by the
homogeneous variant of the relevant sG model~\cite{CKW14})
\begin{equation}\label{phi_wp1}
\phi(0,x)=4 \arctan \left[ \exp \left( \frac{1}{\sqrt{{\cal
F}(x_0)}} \,\, \gamma_0 \, (x-x_0) \right) \right],
\end{equation}
\begin{equation}\label{phi_wp2}
\partial_t \phi(0,x)= -\frac{2 v}{\sqrt{{\cal F}(x_0)}} \,\, \gamma_0 \, \sech \left[  \frac{1}{\sqrt{{\cal
F}(x_0)}} \,\, \gamma_0 \, (x-x_0) \right],
\end{equation}
where $\gamma_0=\frac{1}{\sqrt{1-v^2}}.$ 
It can be seen that an unusual feature in this
connection in the above formulas is the presence of the $\frac{1}{\sqrt{{\cal F}(x_0)}}$ factor. It is standard to assume initial conditions corresponding to ${\cal F}(x_0)=1$, i.e., the homogeneous kink profile. We will discuss the reasons for this choice of initial conditions and their consequences in more detail in Section~\ref{IC}. Moreover, we assume standard, and consistent with the initial conditions,  boundary conditions in the form $\phi(t,d)=2 \pi$ and $\phi(t,-d)=0$, i.e., kink-like Dirichlet boundary conditions.
Nevertheless, this latter choice will not be in
any way critical as our kinks will move far from the
domain boundaries (and, e.g., homoegeneous Neumann
boundary conditions would have worked equally well).
The size of the system on which we are conducting calculations is $2d=200$. Moreover, in the numerical simulations, we mainly used Adams' method \cite{adams}.

\section{Conservative Case}
In the absence of dissipation and forcing, the equation of motion simplifies significantly
\begin{equation}
\label{sine-gordon}
\partial_t^2 \phi - \partial_x (\mathcal{F}(x)\partial_x \phi) + \sin \phi = 0,
\end{equation}
 The Lagrangian for this system takes the form
\begin{equation}\label{L}
L= \int_{-\infty}^{+\infty} dx {\cal L}(\phi) =\int_{-\infty}^{+\infty} dx  \left[ \frac{1}{2}~ (\partial_t
\phi)^2 - \frac{1}{2} ~{\cal F}(x) (\partial_x \phi)^2 -  (1-\cos
\phi) \right].
\end{equation}
By introducing a new field variable $\xi=\xi(t,x)$ through the transformation
\begin{equation}\label{phi}
\phi(t,x)=4 \arctan e^{\xi(t,x)},
\end{equation}
the Lagrangian can be equivalently transformed to the form
\begin{equation}\label{Lxi}
L=4 \int_{-\infty}^{+\infty} dx  \,\, {\rm sech}^2 \xi \left[
\frac{1}{2}\,(\partial_{t} \xi)^2 - \frac{1}{2}\, {\cal
F}(x)(\partial_{x} \xi)^2 -\frac{1}{2} \right].
\end{equation}
Here we have used the identity of $1-\cos \phi = 2 {\rm sech}^2 \xi$  satisfied for the function described by 
Eq.~\eqref{phi}.
So far, the function $\xi=\xi(t,x)$ has been completely arbitrary. At this point we will take its special form, which means choosing a special ansatz. This choice of the analytical form of this function is motivated by the form of the solution obtained in article \cite{Dobrowolski2017} for a constant value of  ${\cal F}$. This ansatz has the form
\begin{equation}\label{xi}
 \xi= \frac{1}{\sqrt{{\cal
F}(x_0(t))}} \,\, \gamma(t) \, (x-x_0(t)).
\end{equation}
Here, in addition to the position $x_0(t)$, we introduce a degree of freedom describing changes in the 
(inverse) width of the kink $\gamma(t)$. This 
variable describes both the dynamic and kinematic effects that cause changes in kink width.
Now the effective Lagrangian is obtained by integrating over the spatial variable as:
\begin{equation}\label{Leff}
L_{eff} = \frac{1}{2}\, M \Dot{x}_0^2 +\frac{1}{2}\, m \Dot{\gamma}^2 - \kappa \Dot{x}_0 \Dot{\gamma} - V.
\end{equation}
The parameters present in the above Lagrangian are defined by the integrals
\begin{equation}
\begin{gathered}
\label{integrals}
    M = 4\int_{-\infty}^{+\infty}dx\sech^2(\xi) \, W(\xi)^2,\\
    m = \frac{4}{\gamma^2}\int_{-\infty}^{+\infty}dx\sech^2(\xi) \, \xi^2,\\
    \kappa = \frac{4}{\gamma} \int_{-\infty}^{+\infty}dx\sech^2(\xi) W(\xi) \, \xi ,\\
    V = 2\int_{-\infty}^{+\infty}dx\sech^2(\xi) \, \left( 1 + \frac{\mathcal{F}(x)}{\mathcal{F}(x_0)}\gamma^2 \right) ,
\end{gathered}
\end{equation}
where the auxiliary function $W$ is the following
\begin{equation} \label{W}
    W(\xi) = \frac{1}{2 {\cal F}(x_0)} ( \partial_{x_0} {\cal F}(x_0)) \, \xi + \frac{1}{\sqrt{{\cal F}(x_0)}} \, \gamma .
\end{equation}
Obviously, all coefficients are functions of the dynamic variables $x_0(t)$ and $\gamma(t)$.
The Euler-Lagrange equations obtained from the effective Lagrangian take the form of
\begin{equation}
\begin{gathered}
\label{2dof_ansatz}
    M\Ddot{x}_0-\kappa\Ddot{\gamma}+\frac{1}{2}(\partial_{x_0}M)\Dot{x}_0^2-\frac{1}{2}(\partial_{x_0}m)\Dot{\gamma}^2-(\partial_{\gamma}\kappa)\Dot{\gamma}^2+(\partial_{\gamma}M)\Dot{\gamma}\Dot{x}_0+\partial_{x_0}V=0,\\
    m\Ddot{\gamma}-\kappa\Ddot{x}_0 +\frac{1}{2}(\partial_{\gamma}m)\Dot{\gamma}^2-\frac{1}{2}(\partial_{\gamma}M)\Dot{x}_0^2-(\partial_{x_0}\kappa)\Dot{x}_0^2+(\partial_{x_0}m)\Dot{x}_0\Dot{\gamma}+\partial_{\gamma}V=0.
\end{gathered}
\end{equation}
The analytical forms of the coefficients appearing in the equations can be found in the Appendix A.

\subsection{Interaction of the kink with the potential barrier}
\begin{figure}[h!]
    \centering
    \subfloat{{\includegraphics[width=7.5cm]{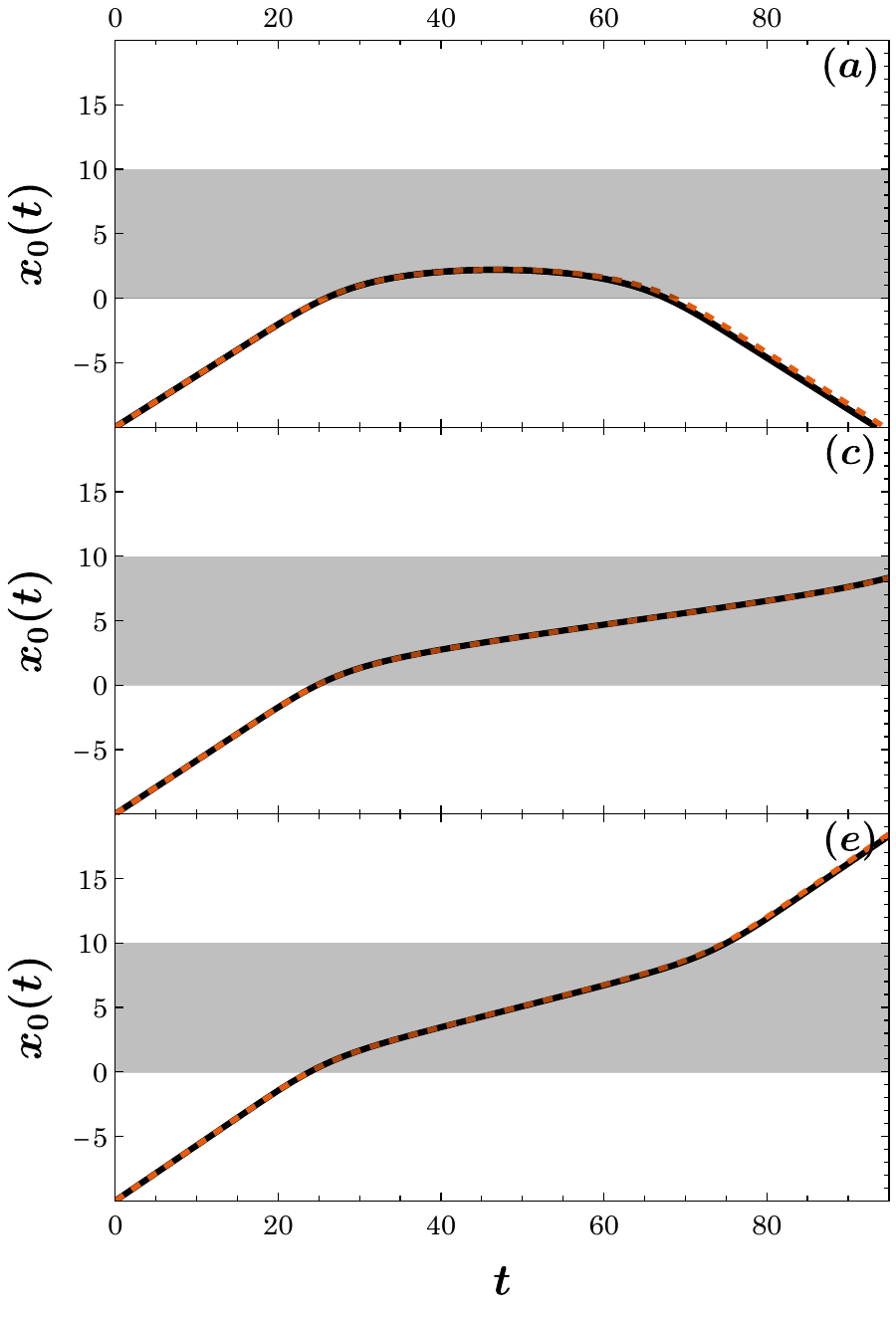}}}
    \quad
    \subfloat{{\includegraphics[width=7.65cm]{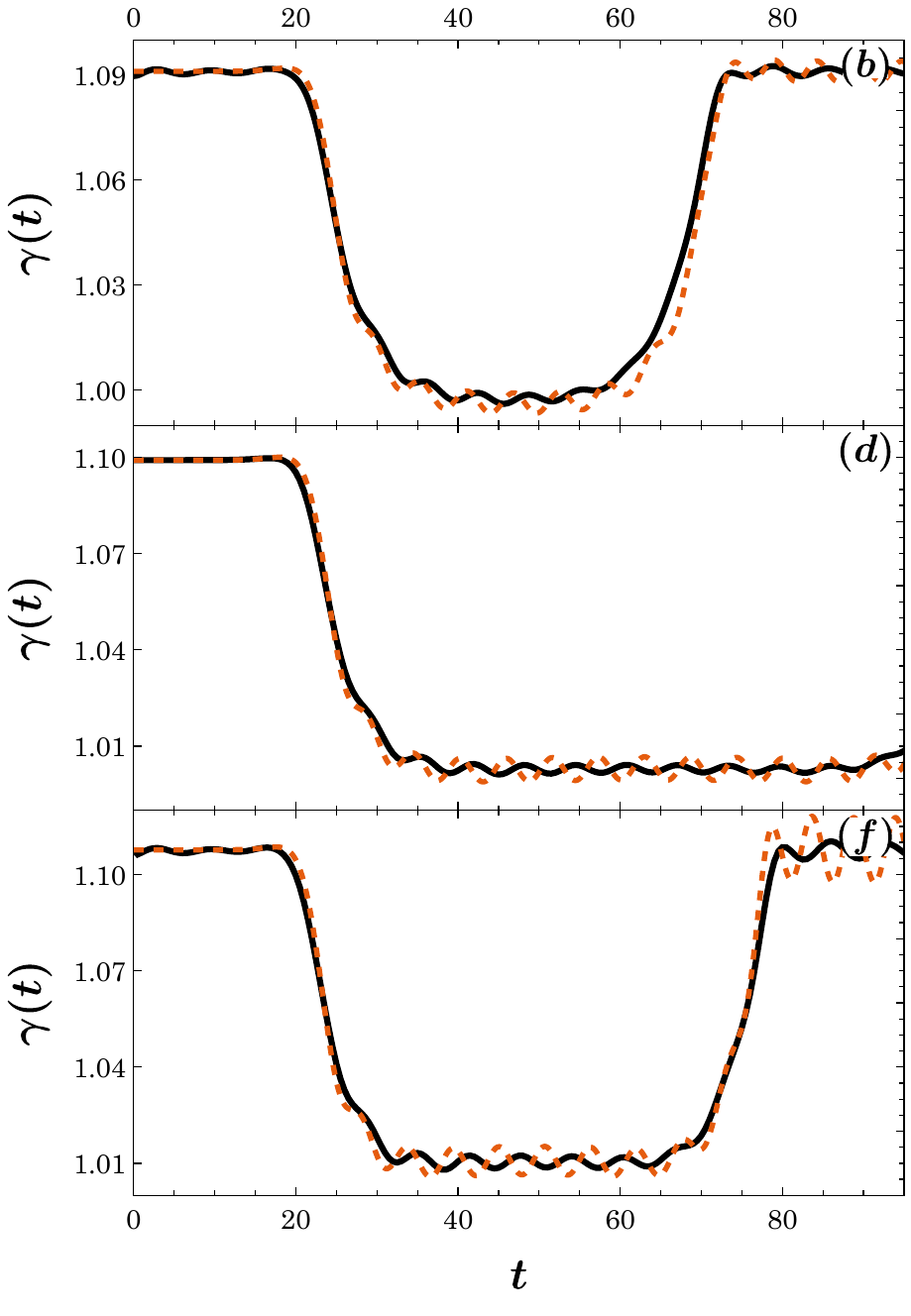}}}
    \caption{Left panels {(a), (c) and (e)}: comparison of the position of the kink  for the solution from the original field model (black line) and the model according to the equation \eqref{2dof_ansatz} (red line). Right panels {(b), (d) and (f)}: analogous comparison for the variable $\gamma(t).$
    The figures are  for the cases of $\varepsilon = 0.1$ and velocities (starting from the top) $0.4$ {(a, b)}, $0.415$ {(c, d)} and $0.43$ {(e, f)}, i.e., subcritical, near-critical and
    super-critical, respectively, with respect to the height of the barrier.}
    \label{fig_01}
\end{figure}

\begin{figure}[h!]
    \centering
    \includegraphics[width=7.5cm]{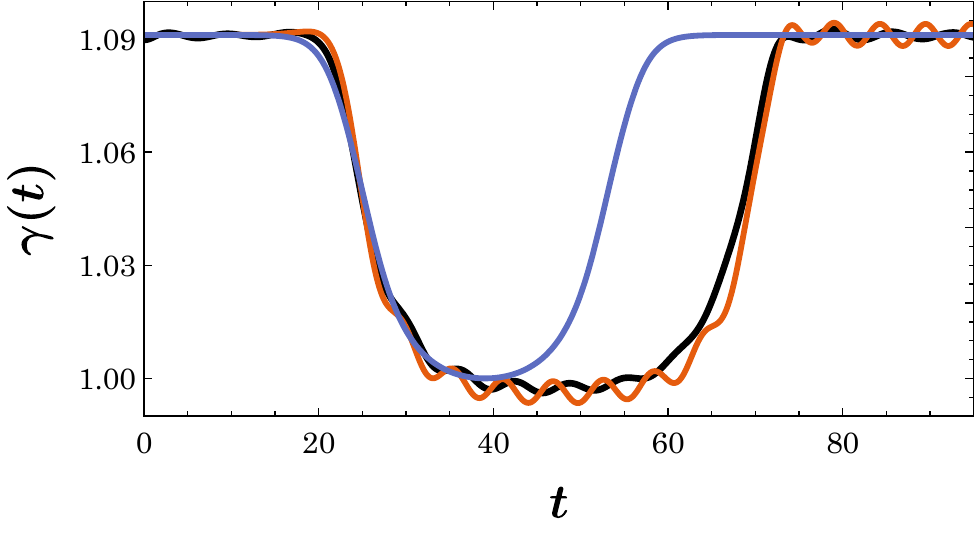}
    \caption{{Comparison of $\gamma(t)$ from the field model (black line), along with the solution based on equation \eqref{2dof_ansatz} (red line), and $\gamma$ in the form of the Lorentz factor without the second degree of freedom (blue line). 
    The latter is clearly less
    satisfactory in capturing
    the PDE dynamics.
    Here $\varepsilon = 0.1$ and $v=0.4$.}}
    \label{fig_02}
\end{figure}

\begin{figure}[h!]
    \centering
    \subfloat{{\includegraphics[width=7.5cm]{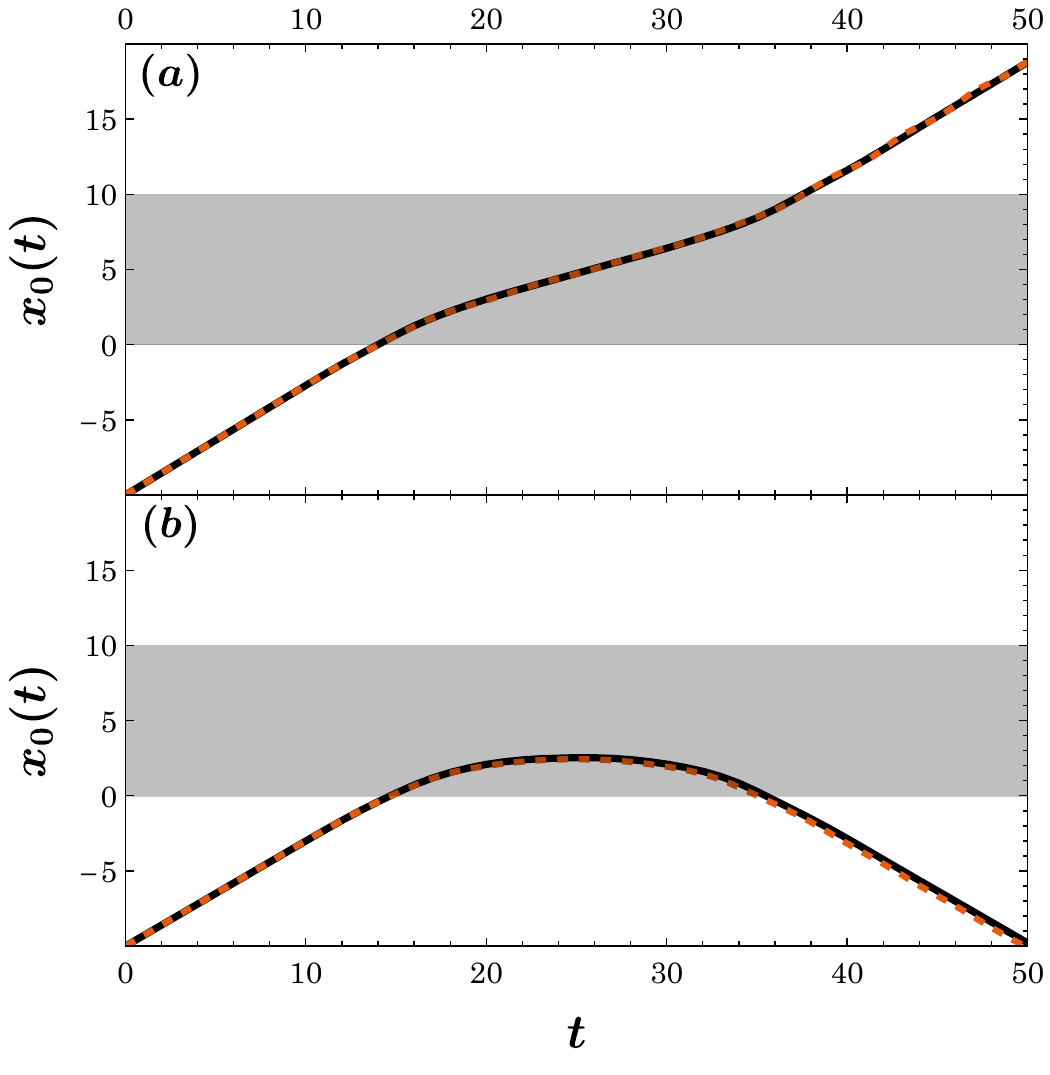}}}
    \quad
    \subfloat{{\includegraphics[width=7.5cm]{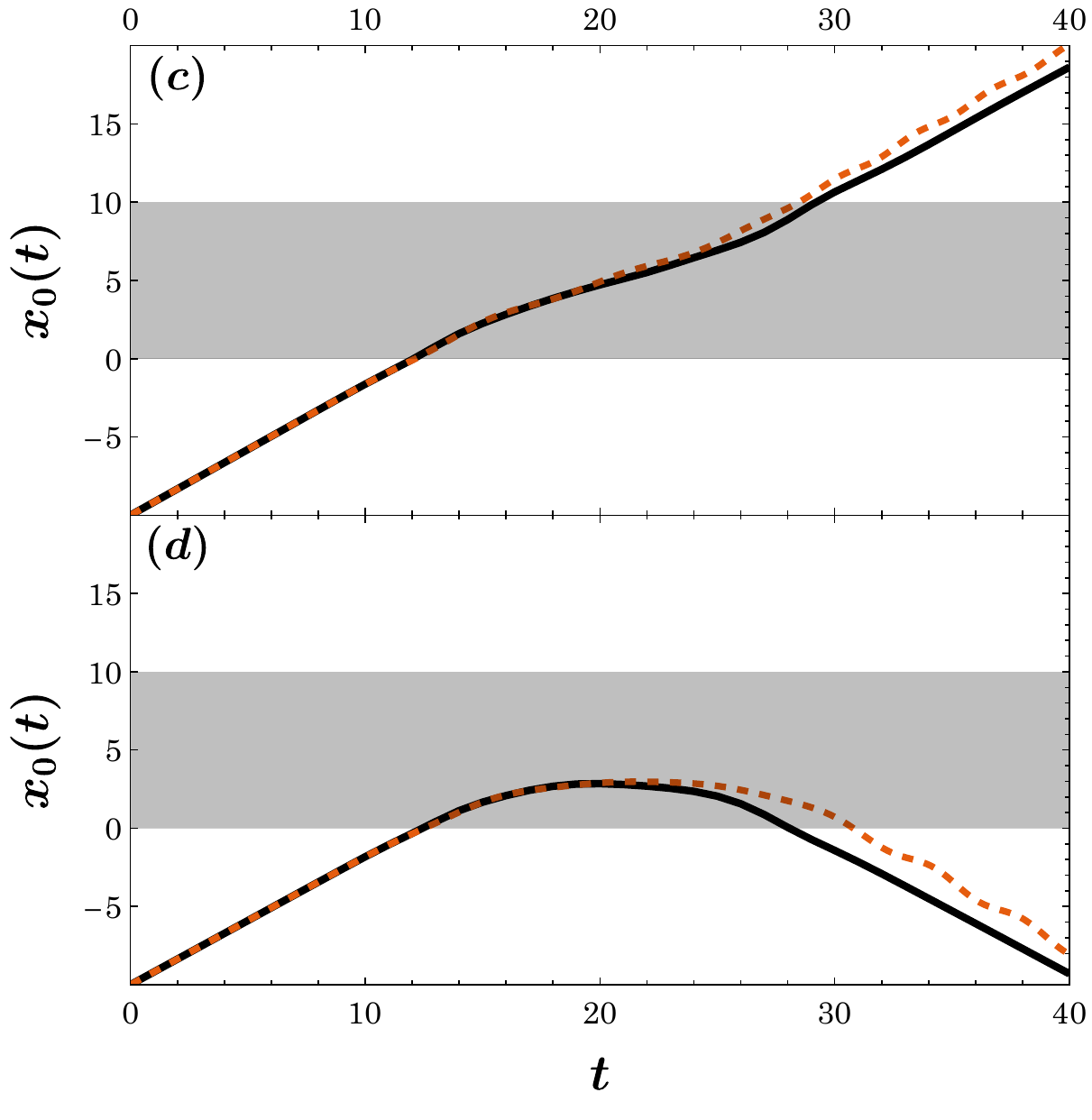}}}
    \caption{{Comparison of the position of the kink  for the solution from the original field model (black line) and the model according to the equation \eqref{2dof_ansatz} (red line). On the left { (a) and (b)} $\varepsilon = 0.5$ and velocities $0.73$ {on (a)} and $0.7$ {on (b)}, and on the right { (c) and (d)} $\varepsilon = 1$ and velocities $0.84$ {on (c)} and $0.82$. {on (d)}}}
    \label{fig_03}
\end{figure}

Although our goal is to study the dynamics of the kink in the perturbed sine-Gordon model in the presence of {\it periodic} inhomogeneities, i.e., an effective
external ``lattice'', prior to that and in order to check the validity of the ansatz, as a first step, we will compare the results with those obtained for the potential barrier in the \cite{Gatlik2023b} article. The analytical form of the $g$-function determining the shape of the barrier is the following
\begin{equation}
    \label{g1}
    g(x) = \tanh(x)-\tanh(x-L) ,
\end{equation}
where $L$ is the thickness of the barrier.
As a reminder, the function ${\cal F}(x)$ appearing in the field equation is given by the formula \eqref{inhomogeneity}.
The results obtained from the field model \eqref{sine-gordon} and the effective model \eqref{2dof_ansatz} are compared in Figure \ref{fig_01}. The left panel (Fig.1.a,c and e) shows the evolution of the variable $x_0$ describing the position of the kink.
 Throughout the article, we identify the location of the kink (in the field model) with the location of the point 
$x$ where, for each $t$,
$\phi(t,x)=\pi$. 
Note that at the level of the field equation, in the examples considered, there is no significant difference between the position of the kink defined by the point $\phi=\pi$ and the position of the center of mass (see Appendix B for details).
The black line represents the result obtained in the field model while the dashed red line represents the result obtained in the effective model. The location of the barrier is represented in the figure by a gray area. The figure assumes an $\varepsilon$ parameter equal to $0.1$,
while the thickness of the barrier is $L=10$. The top left and right figures (Fig.1.a and b) were made for an initial kink velocity of $0.4$. As can be seen (Fig.1.a) this velocity is not sufficient to cross the barrier. The behavior of the variable $\gamma$, responsible for changes in the kink thickness during interaction with the barrier is shown in the upper right figure (Fig.1.b).
In both figures, the comparison of the PDE
model with the effective ODE description is very good. The middle figures (Fig.1.c and d)
show the interaction of the kink with the barrier for speeds close to the critical speed. The initial velocity of the kink is assumed here to be equal to $0.415$. In the case shown in the middle figures, the kink eventually passes over the barrier but spends a long time in the area occupied by the barrier.
The lower figures (Fig.1.e and f) show the passage of the kink over the barrier at speeds clearly exceeding the critical speed. Here, the figures assume an initial velocity equal to $0.43$. In the figures shown, the agreement of the PDE model with the effective particle description for the variable $x_0$ is {\it excellent}. On the other hand, for the effective $\gamma(t)$ we have a slight discrepancy. {Let us notice that the value of the variable $\gamma$ obtained from the PDE equation is obtained in two steps. In the first step, on the basis of the obtained field configuration, the position of the kink $x_0$ is determined. In the second step, for the previously gained $x_0$ by comparing the obtained configuration with the function $\phi(x) = 4 \arctan \left[\exp \left( \frac{1}{\sqrt{{\cal F}(x_0)}} \gamma (x-x_0)\right) \right]$, the value of $\gamma$ is fitted. This two-step approach ensures that the fitting process captures both the spatial localization of the kink and the precise scaling of its width, $\gamma$. It makes the quantity obtained from the field equation practically the same quantity as those obtained from the approximate reduced equations
(assuming that the field retains this
    kink-like form). To compare the results obtained based on the model with two degrees of freedom (ansatz \eqref{xi}) and with one degree of freedom, i.e., when $\gamma$ is determined based only on the velocity of the kink 
(based on the Lorentz factor), we have included a relevant comparison in Figure \ref{fig_02}. The black line in this figure shows the result obtained from the \eqref{sine-gordon-dis} equation, while the blue line shows the result obtained for the model with one degree of freedom and the red line for two degrees of freedom. The figure was obtained for $\varepsilon=0.1$, barrier width $L=10$ and initial velocity $v= 0.4$. It can be seen that the model with two degrees of freedom better describes the width oscillations inside the barrier. Moreover, the exclusion of the second degree of freedom (blue line) results in a shortening of the interaction time between the kink and the barrier, i.e., a
single degree of freedom is
insufficient to correctly capture the details of the interaction 
between the kink and the barrier.
In~\cite{Gatlik2023}, an energy landscape associated with the
relevant interaction can be found. As for the range of effectiveness of the method depending on the value of the $\varepsilon$ parameter, as far as position prediction is concerned, it is very good up to a value of $\varepsilon=0.5$. Above this value, it is also quite good however it gradually decreases in efficiency, up to a value of $\varepsilon=1$, which is shown in Figure \ref{fig_03}. Obviously, the precision when it comes to the variable $\gamma$ for values above $0.1$ gradually degrades
and at $\varepsilon=1$ it is 
far less satisfactory. Yet, this
is certainly understandable under
the present perturbative considerations.}

\subsection{Kink in the presence of periodic inhomogeneity}
\begin{figure}[h!]
    \centering
    \subfloat{{\includegraphics[height=4.5cm]{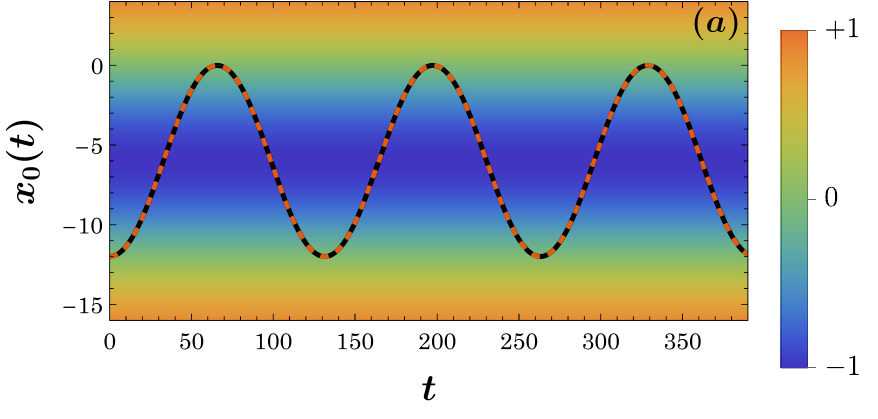}}}
    \quad
    \subfloat{{\includegraphics[height=4.5cm]{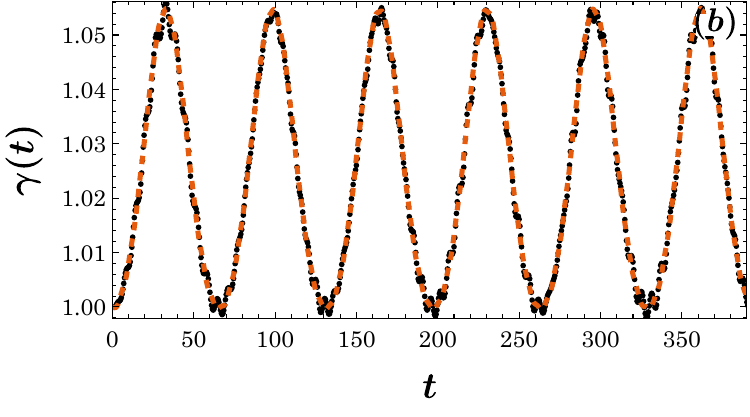}}}
    \quad
    \subfloat{{\includegraphics[height=4.5cm]{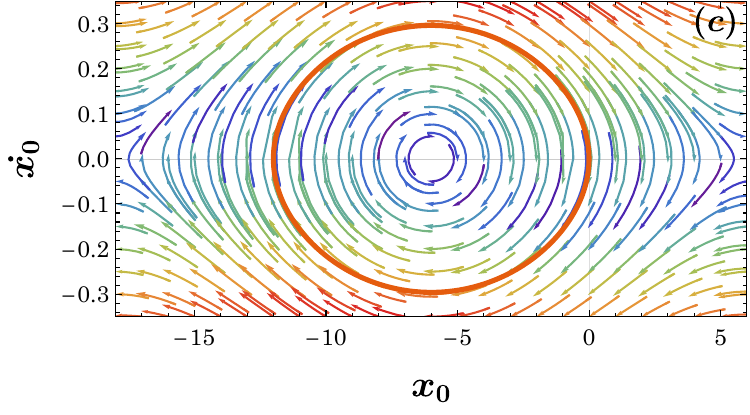}}}
    \caption{{Figure (a)}: comparison of the position of the kink for the solution from the original field model (black line) and the model according to the equation \eqref{2dof_ansatz} (red dashed line) for the periodic potential. The colors in this figure describe the values of the function $g(x)$ (according to the legend to the right). { Figure (b)}: the evolution of the $\gamma(t)$ variable from the PDE model (black points) and effective model (red line). {Figure (c)}: phase portrait of the system in cross section $\gamma=1$ and $\Dot{\gamma}=0$. The red line represents the trajectory shown in the top figure. Here $\varepsilon = 0.1$, and initial velocity is equal $v=0$ and $x_0=-12$.}
    \label{fig_04}
\end{figure}

The second type of inhomogeneity is described by a periodic function
\begin{equation}
\label{sin}
    g(x) = \sin\left(\frac{ \pi }{12} x\right) .
\end{equation}
Note that the "period" of inhomogeneity was chosen so that the kink is much narrower than the period of the inhomogeneity
(with the latter being equal to $24$, while the kink
half-width is $\approx 2.6$).
The first case studied numerically is that of a kink initially located at the first zero of the sine function located to the left of the origin of the coordinate system. 
As before, we take $\varepsilon$ equal to $0.1$. 
Figure \ref{fig_04} shows the movement of the kink between the node of the sine function located at $x=-12$ and  $x=0$. The figure consists of three panels. The top panel { (Fig.4.a)} shows the changes in the position of the kink $x_0$ as a function of time. The colors in the figure represent the values of the function $g(x)$ present in the definition of inhomogeneity \eqref{inhomogeneity}. According to the legend on the right side of the figure, the red color describes values $g$ equal to $+1$ while the smallest values of $-1$ are represented by the blue color. The black line in the figure represents the result obtained based on the PDE model, while the red dashed line was obtained based on the effective derived
ODE model.
It can be seen that for many oscillations the effective model \eqref{2dof_ansatz} maintains very good agreement with the field model \eqref{sine-gordon}. This observation holds true for both the collective variable describing the position of the kink $x_0$ { (Fig.4.a)} and its thickness $\gamma$ shown in the left bottom panel { (Fig.4.b)}. In the bottom left figure, the red points represents the result obtained from the effective model, while the black points were obtained from the field (PDE) model.
As expected, whenever the kink is at the minimum of the function $g(x)$ (the blue area in the top figure),
i.e., it moves the fastest, there is a contraction (the $\gamma$ reaches a maximum value). What's more, the lower right panel  { (Fig.4.c)} 
illustrates the shape of the kink trajectory in phase space. This figure shows the trajectory as a function of the variables $x_0$ and $\Dot{x}_0$ (with $\gamma=1$ and $\Dot{\gamma}=0$ fixed).
The trajectory shown in the top figure { (Fig.4.a)} is represented in the phase portrait { (Fig.4.c)} by the red continuous line. The relevant phase portrait is 
clearly reminiscent of a pendulum-like dynamics.

The next figure, i.e., Figure \ref{fig_05}, shows the transition of the kink between the maximum of the sine function located at $x=-18$ and the adjacent maximum located at $x=6$, i.e., it represents an approximation to the heteroclinic orbit of the relevant effective
particle system. 
As before, $\varepsilon$ is adopted at the level $0.1$. This time the initial velocity of the kink is
very weakly non-zero (i.e., $0.01$) and is directed toward the adjacent potential energy maximum.
The symbols utilized in Figure \ref{fig_05} are identical to those used in Figure \ref{fig_04}.
The top panel { (Fig.5.a)} shows a comparison of the results of the PDE model and of the effective model on the background showing the values of the function $g(x)$.
The black line in this figure shows the trajectory obtained from the field model while the red dashed line represents the result from the effective model. 
We can observe a remarkable coincidence between
the two results.
The left bottom panel  of this figure { (Fig.5.b)} compares the trajectories for the $\gamma$ variable obtained on the ground of the effective model (red dots) with analogous results obtained based on the PDE model (black dots).
 From the figure, it can be seen that the largest kink contraction (increase of $\gamma$) occurs when the kink moves the fastest, passing through the minimum of the $g(x)$ function. The trajectory shown in the top panel corresponds in phase portrait { (Fig.5.c)} to the red line connecting the two saddle points. Obviously, both Figure \ref{fig_04} and Figure \ref{fig_05} show the behavior of the kink in the absence of dissipation and bias current.
\begin{figure}[h!]
    \centering
    \subfloat{{\includegraphics[height=4.5cm]{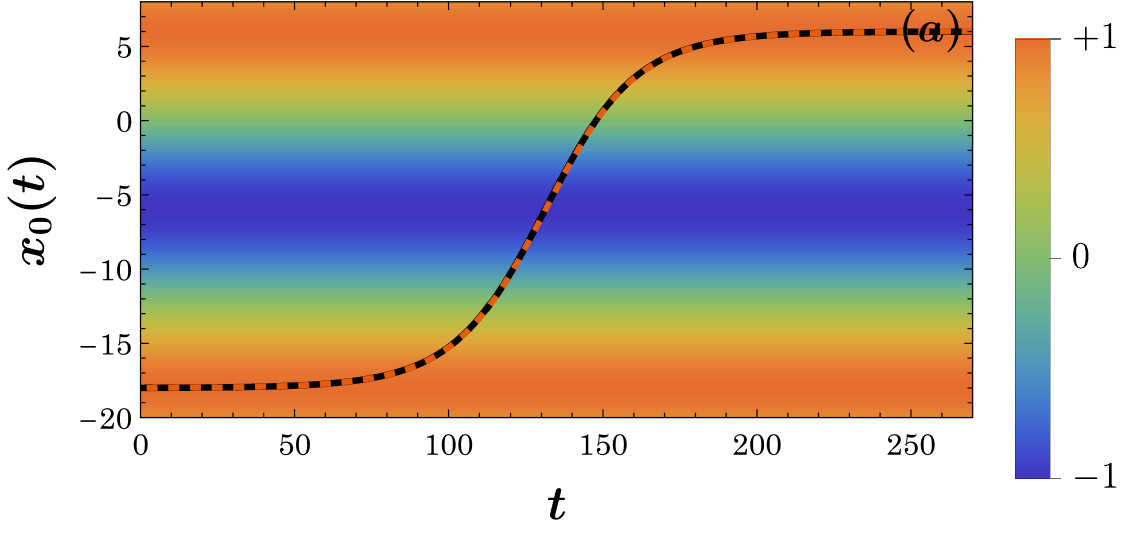}}}
    \quad
    \subfloat{{\includegraphics[height=4.5cm]{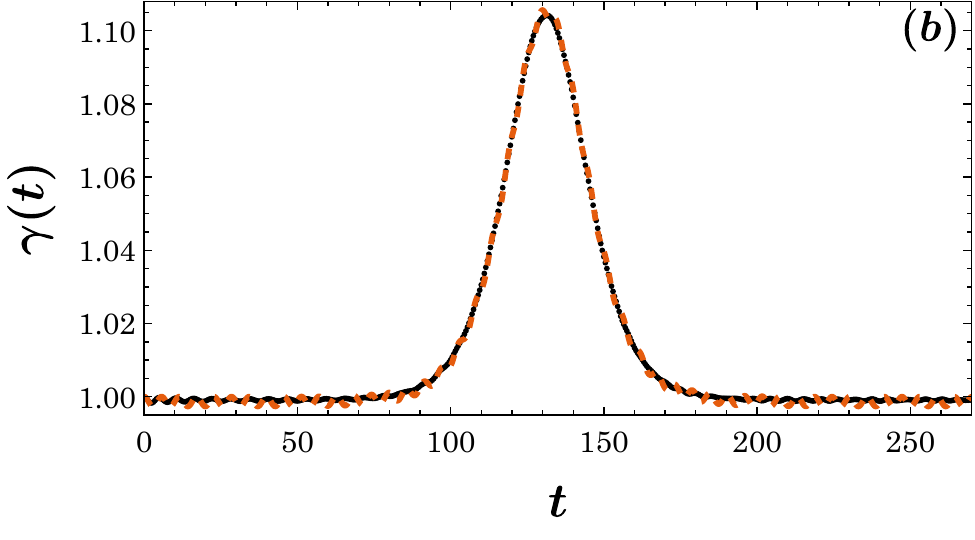}}}
    \quad
    \subfloat{{\includegraphics[height=4.5cm]{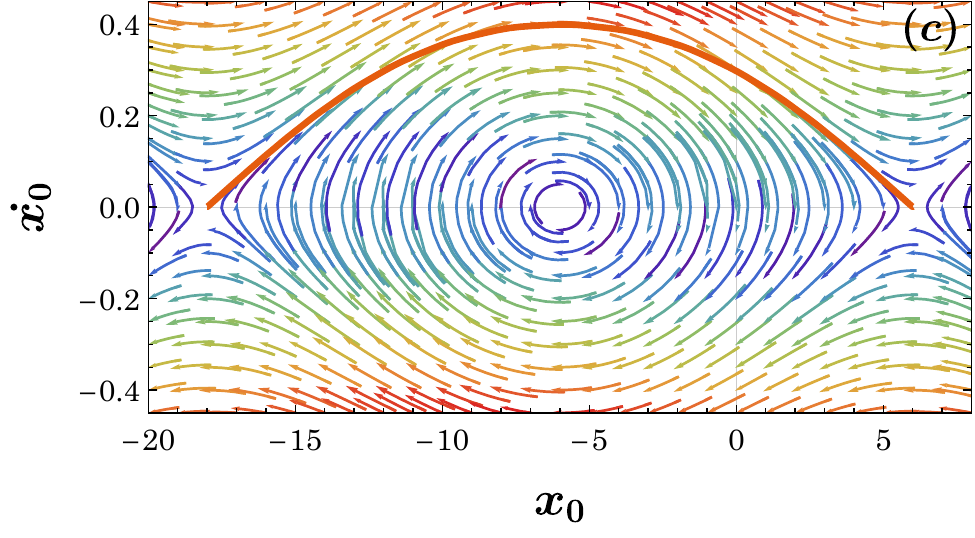}}}
    \caption{{Figure (a)}: comparison of the $x_0$ variable  for the solution from the  field model (black line) and the effective model (red dashed line). The colors in this figure describe the values of the function $g(x)$. {Figure (b)}: the evolution of the $\gamma(t)$ variable from the field model (black points) and effective model (red line). {Figure (c)}: phase portrait of the system (for $\gamma=1$ and $\Dot{\gamma}=0$). The red line represents the trajectory shown in the top figure. Here $\varepsilon = 0.1$, initial velocity is equal $v=0.001$ and initial position $x_0=-18$.}
    \label{fig_05}
\end{figure}

\section{Comment on the choice of initial conditions}
\label{IC}

Here we would like to justify the choice of initial conditions described by equations \eqref{phi_wp1} and \eqref{phi_wp2}. In order for the evolution to proceed smoothly, we should choose initial conditions as close as possible to the configuration  that minimizes energy and corresponds to the solution of the equation of motion.  On the other hand, we know from~\cite{Dobrowolski2017} that in the considered system, there exists a static solution for a fixed value of ${\cal F}.$ This solution contains a dependence on the {  $\frac{1}{\sqrt{{\cal F}(x_0)}}  (x-x_0)$ } variable, not on the $(x-x_0)$ variable. Our goal is to select a kink configuration that approximates the true solution of equation of motion in the best possible way, at least in the initial moments of evolution. The stationary kink that best approximates (at least initially) the solution of the field equation takes the form of
\begin{equation}
    \label{stacionary}
    \phi(t,x)=4 \arctan \left[ \exp \left( \frac{1}{\sqrt{{\cal
F}(x_0)}} \,\, \gamma_0 \, (x-x_0-vt). \right) \right],
\end{equation}
where $\gamma_0 = \frac{1}{\sqrt{1-v^2}}$ and $v$ is constant velocity.
We additionally include a $\gamma_0$ factor in the initial condition because we are interested in a fully relativistic stationary kink configuration in order to be able to describe
the kink motion ({ including} the contraction of the kink
width). This setup at time $t=0$ allows one to define the initial conditions given by the formulas \eqref{phi_wp1} and \eqref{phi_wp2}.
The smooth evolution, capturing not only the
kink motional degree of freedom but also the
effective kink width evolution, is particularly evident in the left panel of the Figure \ref{fig_05}, which describes the dependence of the variable $\gamma$ on time. This result was obtained with the described initial conditions.

\begin{figure}[ht]
    \centering
    \includegraphics[height=4.5cm]{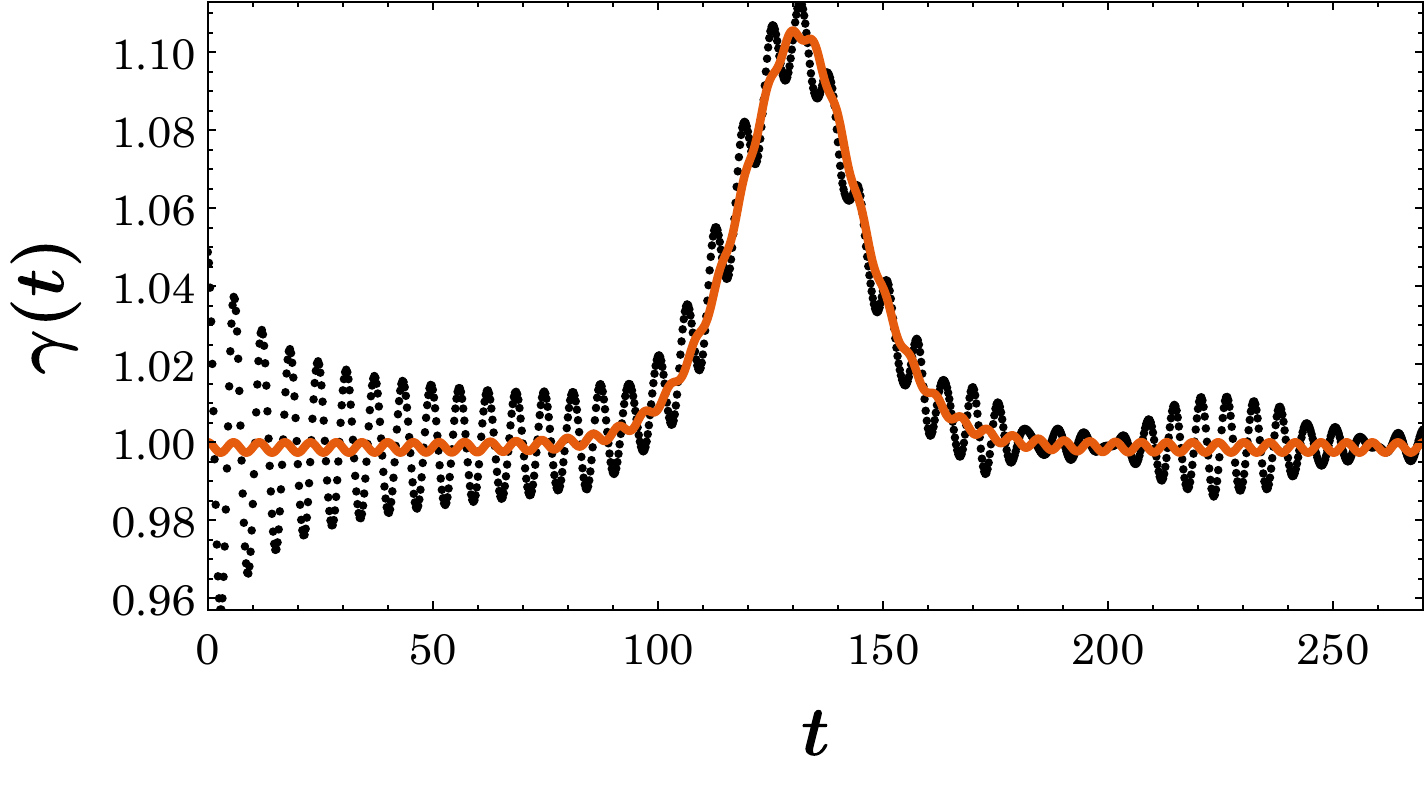}
    \caption{The kink width obtained from the field equation (black dots) at standard initial conditions (i.e., for ${\cal F}=1$) shows significant oscillations especially at the beginning of the evolution. The $\gamma(t)$ variable is compared with the result of the effective model. The parameters in the figure are $\varepsilon = 0.1$, $v=0.001$ and  $x_0=-18$.}
    \label{fig_06}
\end{figure}

For comparison, in Figure \ref{fig_06} we show the course of evolution in the case of standard initial data, i.e., data with ${\cal F}=1$. It can be seen that due to the non-matching of the data with the true solution, there are rapid oscillations at the beginning of the evolution. 
Due to the lack of dissipation, this excess energy present in the system also appears at later stages of evolution. The figure similarly to the previous one shows the behavior of the $\gamma$ variable for the case when kink goes from one maximum of the inhomogeneity described by the \eqref{sin} equation to the neighboring maximum. 
The figure was obtained for parameter values identical to those in the previous figure, i.e.  $\varepsilon = 0.1$, initial velocity is equal to $v=0.001$ and initial position $x_0=-18$, but only varying the initial
condition as described above.

\section{Moduli space of the system}
\begin{figure}[ht]
    \centering
    \subfloat{{\includegraphics[height=4.5cm]{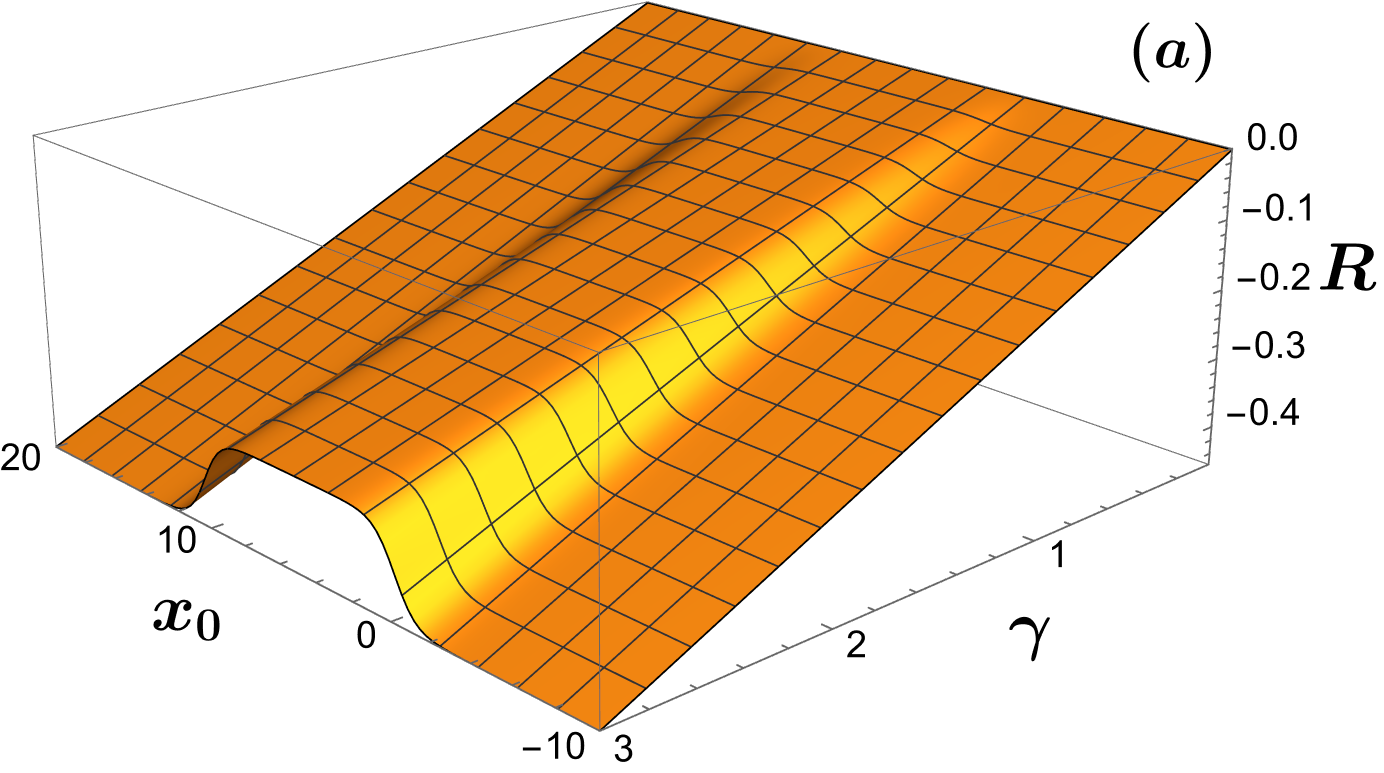}}}
    \quad
    \subfloat{{\includegraphics[height=4.5cm]{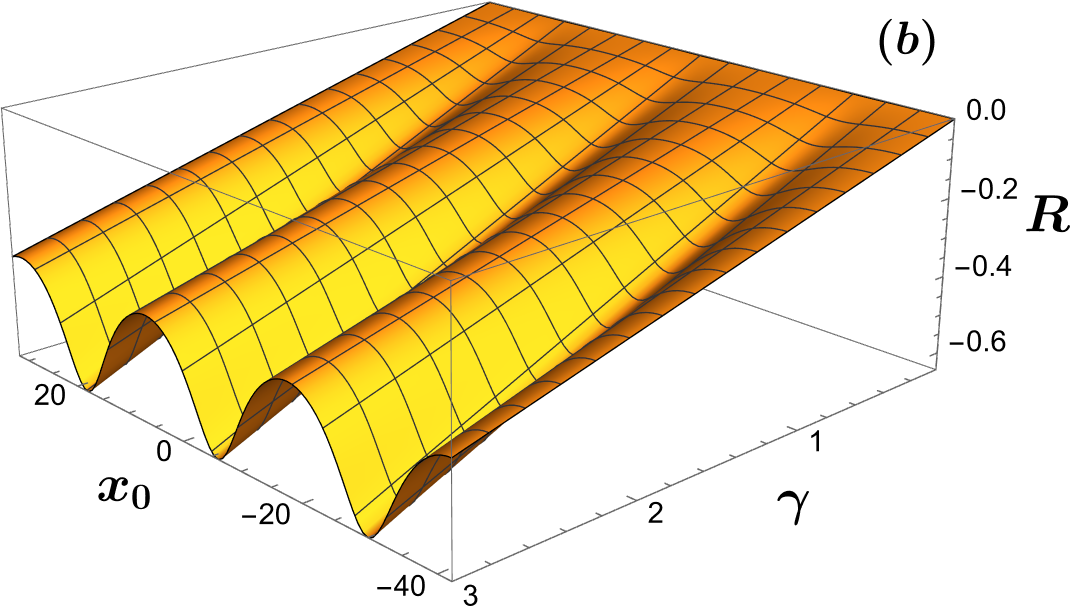}}}
    \caption{Ricci scalar for inhomogeneity in the form of a barrier {(a)} with $L=10$ and periodic inhomogeneity {(b)}. In both cases  $\varepsilon = 0.5$. }
    \label{fig_07}
\end{figure}

The quadratic form of the term describing the kinetic energy in the Lagrangian \eqref{Leff} naturally defines the metric of the moduli space in the parameterization given by the collective variables $x_0$ and $\gamma$. This metric is defined by the functions $M=M(x_0,\gamma)$, $m=m(x_0,\gamma)$ and $\kappa=\kappa(x_0,\gamma)$ given in the Appendix A:
\begin{equation}
    \label{gg0}
    g_{x_0 x_0} = M, \,\,\,\,\,  g_{\gamma \gamma} = m, \,\,\,\,\, g_{x_0 \gamma} = g_{\gamma x_0} = -\kappa .
\end{equation}
In order to describe the properties of the modular space of the system under study, we will not refer to the forms of the components of the metric, which themselves are closely related to the way the parameterization is chosen, but we will refer to the invariant of the parameterization, which is the curvature scalar.  
The Ricci scalar ${ R}$ in the cases we are considering is uniquely defined by, among other things, the form of the function ${\cal F}(x_0)$
{
\begin{equation}
\label{RF}
{\mathbf R} = \frac{36 \gamma^3 \sqrt{{\cal F}(x_0)} \left( -4 \pi^2 \gamma
(\gamma-2) {\cal F}(x_0) \partial^{\,2}_{x_0} {\cal F}(x_0) +
\pi^2 ( \gamma (2 \gamma-7) +4) (\partial_{x_0}{\cal F}(x_0))^2
-48 \gamma^3 {\cal F}(x_0) \right)}{\left(48 \pi \gamma^3 {\cal
F}(x_0) + \pi^3 (\gamma -1)
\partial^{\,2}_{x_0} {\cal F}(x_0) \right)^2} .
\end{equation}
}
Figure \ref{fig_07} shows the values of the curvature scalar as a function of the $x_0$ and $\gamma$ quantities. The left panel {(Fig.7.a)} of this figure describes the curvature of the moduli space for inhomogeneities of barrier form \eqref{g1} of width $L=10$. The right panel {(Fig.7.b)}, in turn, shows the curvature of the space when the inhomogeneity is of periodic form \eqref{sin}. In both cases, $\varepsilon=0.5$ is assumed. It can be seen that the form of the curvature is closely related to the shape of the inhomogeneity. In both cases, the space is free of singularities. 

\begin{figure}[ht]
    \centering
    \includegraphics[height=4.5cm]{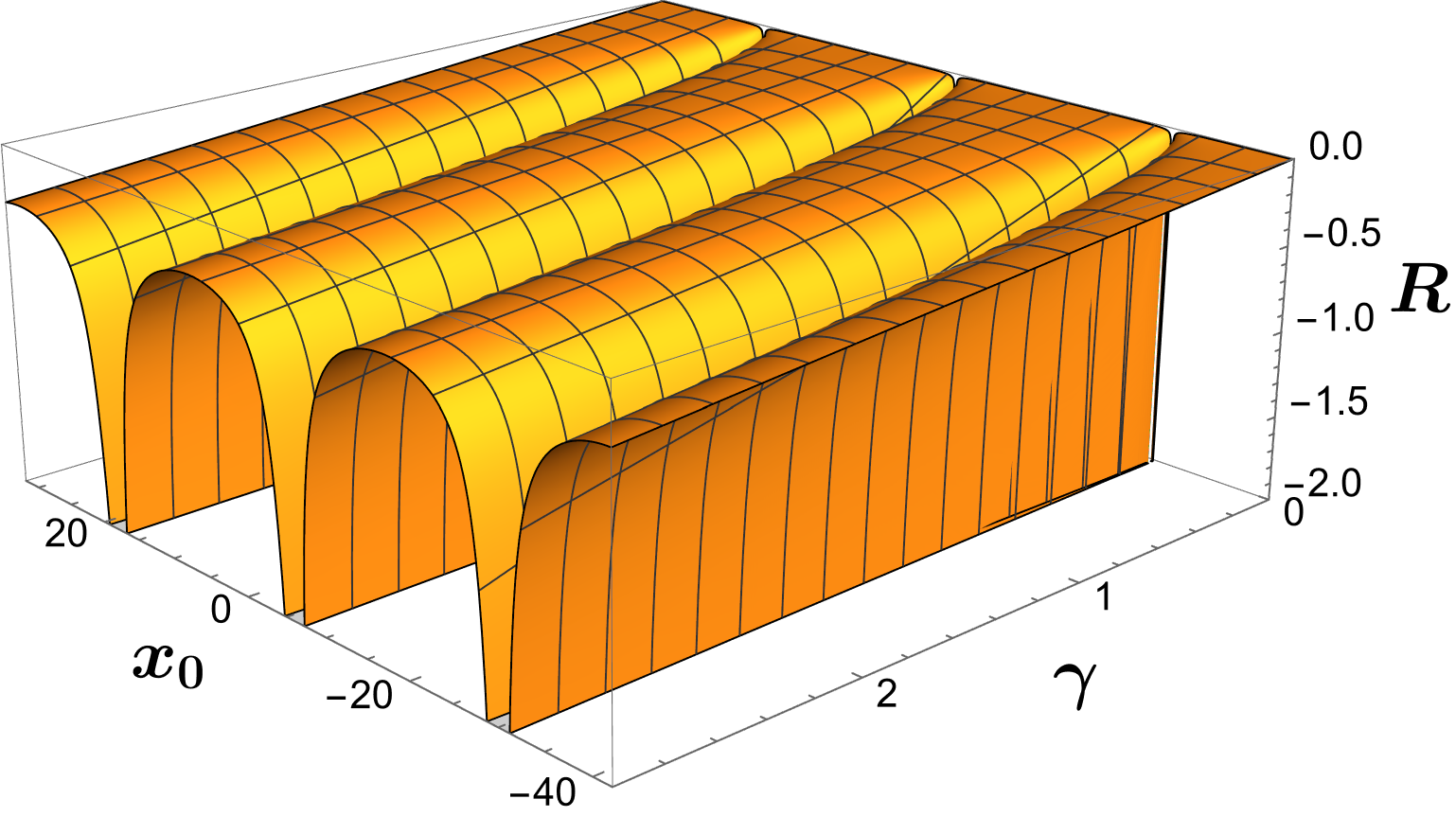}
    \caption{Ricci scalar for periodic heterogeneity at  $\varepsilon = 1$. Singularities occur periodically along  lines $x_0=-30$, $x_0=-6$ and $x_0=18$.}
    \label{fig_08}
\end{figure}

Naturally, a singularity in moduli space can be easily generated by choosing the appropriate form of the function ${\cal F}(x_0)$ (here defined by the  function $g(x_0)$). In the cases we are considering, the singularity appears for the periodic function \eqref{sin} if only we take a sufficiently large value of the parameter $\varepsilon$.  Figure \ref{fig_08} shows the singularities that appear in moduli space (for $\varepsilon=1$) along several lines $x_0=-30$, $x_0=-6$ and $x_0=18$. These singularities correspond to the minima of the sine function. The singularities that appear here are closely related to the change in the nature of the equation \eqref{sine-gordon}. The singularities shown in this figure are responsible for the zeroing of the ${\cal F}$ function, and thus cause the term containing derivatives after the spatial variable $x$ to disappear in the equation \eqref{sine-gordon}. In this case the equation becomes an ordinary differential equation. Obviously, in this case, our ansatz \eqref{xi} also loses its meaning. 

\section{Dynamics in the presence of dissipation}
In the description of the system containing dissipation and external forcing, and given by equation of motion \eqref{sine-gordon-dis}, we will use the method described in \cite{Galley2013,Kevrekidis2014}. According to this approach, the physical variables are duplicated.  The nonconservative Lagrangian  
${\cal L}_N$, describing the system under study, contains conservative parts for first ${\cal L}(\phi_1)$ and the second variable ${\cal L}(\phi_2)$, as well as a part that reproduces the non-conservative terms ${\cal R}$ in the field equation
\begin{equation}
    \label{nonconservative}
    {\cal L}_N = {\cal L}(\phi_1) - {\cal L}(\phi_2) + {\cal R} ,
\end{equation}
where the analytical form of the Lagrangian density for $\phi_1$ and $\phi_2$ is for each of the two variables the same as for the field variable $\phi$ in the equation \eqref{L}. 
{ The need to formulate a variational principle based on the action formed on the basis of this type of Lagrangian was raised in the article \cite{Galley2013}.  It was pointed out 
therein that Hamilton's principle is formulated as a boundary value problem in time (between initial and final times). On the other hand the equations of motion are solved as an initial value problem. Moreover the standard formulation of the variational principle leaves no room for describing processes that are not 
time translation invariant (i.e., non-conservative). In order to formulate the variational principle that is compatible with initial value problem and which enables the description of processes that are non-conservative,  the number of degrees of freedom was doubled, allowing for the action of non-conservative
forces between the initial and final time. When the Lagrangian is properly (as above) formed as shown in~\cite{Galley2013}, the vanishing of the variation of the action at the final moment is obtained by equating the arbitrary variances of the two variables. All that remains is to require the vanishing of the variances of both variables at the initial moment. In this way, a variational problem defined by assigning the values of the fields $\phi_1$ and $\phi_2$ only at the initial moment was obtained.
The variables $\phi_1$ and $\phi_2$ used in this procedure are auxiliary variables, however if $\phi_1 \rightarrow \phi_2$ 
after the extremization procedure then the limit is identified with the physical variable $\phi$.
In other words,
the process of identifying the two trajectories and that of performing the
extremization (of the action) do not commute and the trick developed in~\cite{Galley2013}, by first extremizing and
then using $\phi_1 \rightarrow \phi_2$,
enabled the incorporation of non-conservative forces in a Lagrangian formulation.}

A particularly convenient way to write the equations of motion in this approach is one that separates the conservative part from the non-conservative part. In this notation, the equation of motion is of the form
\begin{equation}
    \label{eq-nonconservative}
    \partial_{\mu} \left( \frac{\partial {\cal L}}{\partial (\partial_{\mu} \phi)}\right) - \frac{\partial {\cal L}}{\partial \phi} = \left[\frac{\partial {\cal R}}{\partial \phi_{-}} - \partial_{\mu} \left( \frac{\partial {\cal R}}{\partial (\partial_{\mu} \phi_{-})}\right)\right]_{PL},
\end{equation}
where the usual summation convention has been assumed.
Here, $\phi_{-}$ as well as $\phi_{+}$ are variables related to the original variables $\phi_1$ and $\phi_2$ through $\phi_1=\phi_{+}+ \frac{1}{2}\phi_{-}$ and $\phi_2=\phi_{+}- \frac{1}{2}\phi_{-}$ relationships.
In addition, the inscription PL denotes the so-called physical limit, in which $\phi_{+}$ becomes a physical variable $\phi_{+}=\phi$ and $\phi_{-}$ disappears from the description, i.e., $\phi_{-}=0$. The index $\mu$ enumerates the space-time variables $x^{\mu}=(x^0,x^1)=(t,x)$. In the case of Lagrangian density \eqref{L}, the left-hand side of this equation takes the form
\begin{equation}
\label{eq-nonconservative2}
    \partial_t^2 \phi - \partial_x (\mathcal{F}(x)\partial_x \phi) + \sin \phi = \left[\frac{\partial {\cal R}} {\partial \phi_{-}} - \partial_{\mu} \left( \frac{\partial {\cal R}}{\partial (\partial_{\mu} \phi_{-})}\right)\right]_{PL} .
\end{equation}
It turns out that we can reproduce equation \eqref{sine-gordon-dis} by taking the following form of the non-conservative term
\begin{equation}
   {\cal R} = - \Gamma \phi_{-} - \alpha \phi_{-} \partial_t \phi_{+} .
\end{equation}
The effective description in the case under consideration is obtained by inserting the ansatz defined by equations \eqref{xi} and \eqref{phi} into the expression for ${\cal R}$, and then performing integration over the spatial variable i.e. $R_{eff} = \int_{-\infty}^{+\infty} dx {\cal R}$.
The equations of motion at the effective ODE level are then 
obtained as follows
\begin{equation}
    \label{eff-eq1}
    \frac{d}{d t }\left(\frac{\partial L_{eff}}{\partial \Dot{x}_0} \right) - \frac{\partial L_{eff}}{\partial {x}_0} = \left[ \frac{\partial R_{eff}}{\partial {x}_{-}} - 
    \frac{d}{d t }\left(\frac{\partial R_{eff}}{\partial \Dot{x}_{-}} \right) \right]_{PL} ,
\end{equation}
\begin{equation}
    \label{eff-eq2}
    \frac{d}{d t }\left(\frac{\partial L_{eff}}{\partial \Dot{\gamma}} \right) - \frac{\partial L_{eff}}{\partial {\gamma}} = \left[ \frac{\partial R_{eff}}{\partial {\gamma}_{-}} - 
    \frac{d}{d t }\left(\frac{\partial R_{eff}}{\partial \Dot{\gamma}_{-}} \right) \right]_{PL} ,
\end{equation}
where $L_{eff}$ is defined by the equation \eqref{Leff}.
I.e., the Lagrangian portions of the equations 
obtained previously from Eq.~\eqref{Leff} are now
augmented due the presence of the driving and damping terms of the right
hand side.
This time the physical limit means adopting a limit in which $x_{-}=0$, $\gamma_{-}=0$, $x_{+}=x_0$ and $\gamma_{+}=\gamma$. 
The relationships for effective variables are analogous to those for field variables (i.e.,
$x_1=x_+ + \frac{1}{2} x_-$, $x_2=x_+ - \frac{1}{2} x_-$ and $\gamma_1=\gamma_+ + \frac{1}{2} \gamma_-$, $\gamma_2=\gamma_+ - \frac{1}{2} \gamma_-$).
The final form of the equations of motion is as follows
\begin{equation}
\begin{gathered}
\label{2dof_ansatz-dyss}
    M\Ddot{x}_0-\kappa\Ddot{\gamma}+\frac{1}{2}(\partial_{x_0}M)\Dot{x}_0^2-\frac{1}{2}(\partial_{x_0}m)\Dot{\gamma}^2-(\partial_{\gamma}\kappa)\Dot{\gamma}^2+(\partial_{\gamma}M)\Dot{\gamma}\Dot{x}_0+\partial_{x_0}V= \\  2 \pi \Gamma - \alpha \frac{\pi^2}{6 \gamma} \sqrt{{\cal F}(x_0)} \left( \frac{\partial_{x_0} {\cal F}(x_0)}{{\cal F}(x_0)}\right)^2 \Dot{x}_0 - 8 \alpha \frac{\gamma}{\sqrt{{\cal F}(x_0)}} \Dot{x}_0 + \alpha \frac{\pi^2}{3 \gamma^2} \frac{\partial_{x_0} {\cal F}(x_0)}{\sqrt{{\cal F}(x_0)}} \Dot{\gamma}
    ,\\
    m\Ddot{\gamma}-\kappa\Ddot{x}_0 +\frac{1}{2}(\partial_{\gamma}m)\Dot{\gamma}^2-\frac{1}{2}(\partial_{\gamma}M)\Dot{x}_0^2-(\partial_{x_0}\kappa)\Dot{x}_0^2+(\partial_{x_0}m)\Dot{x}_0\Dot{\gamma}+\partial_{\gamma}V= \alpha \frac{\pi^2}{3 \gamma^2} \frac{\partial_{x_0} {\cal F}(x_0)}{\sqrt{{\cal F}(x_0)}} \Dot{x}_0 - \alpha \frac{2 \pi^2 }{ 3 \gamma^3} \sqrt{{\cal F}(x_0)} \Dot{\gamma}.
\end{gathered}
\end{equation}
The parameters appearing in the above equations were defined by the formulas in equations \eqref{integrals}, and their final forms are given in equations \eqref{integrals-fin}.

\subsection{Kink-barrier interaction in the presence of bias current and dissipation}
Within the palette of damped-driven settings, the first situation to be studied is the interaction of the kink with the potential barrier \eqref{g1} in the presence of a constant bias current {{and dissipation}. } 
In the case examined in this section, the bias current was selected so that the kink velocity does not exceed a critical value. We remember that due to the presence of dissipation, the kink does not accelerate to velocities close to the Swihart velocity \cite{Scott1978}, but its velocity can reach at most a certain stationary value. 
In a homogeneous system, the value of this velocity is determined by the magnitude of the bias current and the damping factor. If we had chosen the initial velocity of the kink higher, it would have been reduced to a stationary value due to dissipation. In the opposite case, i.e., for an initial velocity of the kink smaller we can observe its increase to the stationary velocity which would be due to the existence of a forcing in the form of a bias current. In fact, to simplify the process of interaction of the kink with the barrier, the initial value of the velocity was chosen equal to the stationary velocity (as in the article 
\cite{McLaughlin1978})
\begin{equation}
    \label{u}
    v_s=\frac{1}{\sqrt{1+ \left( \frac{4 \alpha}{\pi \Gamma}\right)^2}} .
\end{equation}
That is to say, we ``align'' the dynamics with those of the
system's attractor, avoiding the transient stage (in which the
speed adapts itself to $v_s$).
The course of the process of interaction of the kink with the barrier is quite simple. Initially, the kink moves with stationary velocity in the direction of the barrier. After hitting the barrier, the kink is reflected towards the initial position.  The presence of an external forcing directed toward the barrier causes the kink to turn around and move toward the barrier again. Subsequent bounces follow in the same manner. The presence of dissipation makes the kink move less and less away from the barrier after subsequent bounces. The final
outcome of this process is to stabilize the position of the kink at a certain (fixed) distance from the barrier. The interaction process is shown in Figure \ref{fig_10}. The left panel of the figure shows the time dependence of the position of the kink. The figure compares the results of the field model \eqref{sine-gordon-dis} (black line) and the effective model \eqref{2dof_ansatz-dyss} (dashed red line).
From a dynamical systems perspective, the corresponding description
is that the concurrent presence of the driving and damping
along with the effective potential creates an {\it additional}
fixed point (around which the damped oscillation observed
in the figure takes place). This new fixed point is
a {\it stable spiral}, hence the corresponding complex
eigenvalue thereof characterizes the oscillatory attraction to
this equilibrium; see also the discussion below.

In the figure, the location of the barrier is marked by a gray area. The right panel contains a similar comparison but it is for the $\gamma$ variable. 
It is remarkable that despite the rather complex dependence
of $\gamma(t)$, the effective corresponding ODE very accurately
captures the relevant dynamics. In the case of the position of
the center $x_0(t)$, the effective particle description is
essentially indistinguishable from the actual PDE dynamics.
The parameters describing the barrier in the figure are $\varepsilon=0.1$ and barrier width $L=10$. The parameter describing the dissipation is equal to $\alpha=0.01$, while the bias current is $\Gamma=0.0045$. According to the \eqref{u} formula, the initial velocity of the kink was chosen to be $v_s \approx 0.33$.

Estimation of the equilibrium position of kink can be made based on the equations~\eqref{2dof_ansatz-dyss}. At equilibrium,
we have that $\Dot{x}_0=0$ and $\Dot{\gamma}=0$. { 
Here we take advantage of the fact that the static solutions of the model without and with dissipation are identical.}
The potential described by the equation~\eqref{integrals-fin} for an inhomogeneity in the form of the barrier \eqref{g1} is the following 
\begin{equation}
    \label{V-barrier}
    V(x_0,\gamma) = \frac{4 \sqrt{{\cal F}(x_0)}}{\gamma} + \frac{4 \gamma}{\sqrt{{\cal F}(x_0)}} + \frac{2 \gamma^2}{
{\cal F}(x_0)} \,\, \varepsilon \int_{-\infty}^{+\infty} dx \sech^2\left(\frac{1}{\sqrt{{\cal F}(x_0)}} \gamma (x-x_0) \right) \left( \tanh(x)-\tanh(x-L) \right) ,
\end{equation}
 Moreover, assuming the Lorentz factor equal to $\gamma \approx 1$ (i.e.,
 the static case), we obtain a nonlinear algebraic equation for the equilibrium position of $x_0$
\begin{equation}
\label{dV}
    \partial_{x_0} V = 2 \pi \Gamma .
\end{equation}
The solution of this equation,  stays in good agreement with the result presented in Figure \ref{fig_10}, i.e. the equilibrium value of the position is $x_0=-2.62$. 
A comparison of this result with the result of the field equation simulation, shown in {(Fig.9.a)}, is represented by the horizontal blue line. { The value of $\gamma \approx 1$ was confirmed by numerical simulations in the field model. Note that in Figures \ref{fig_10} and \ref{fig_11} the value appears to be slightly larger than one. The explanation for this fact is the presence of a force that not only caused the kink to move towards the barrier, but also causes an effect 
 of slightly squeezing the kink at equilibrium. This force presses the kink against the barrier, causing its thickness to decrease.}
{ Moreover, we can estimate the equilibrium kink width based on second equation of the system ~\eqref{2dof_ansatz-dyss}, which in the static case (i.e., assuming that all time derivatives are zero) takes the form:
\begin{equation}
\label{dgV}
    \partial_{\gamma} V = 0 .
\end{equation}
Based on this equation, we can determine the 
equilibrium value of kink thickness. For example, for $\varepsilon=0.5$ from the last equation, we obtain a $\gamma$ value of $1.004$. This value is very close to the $\gamma=1.003$ value obtained based on the field model \eqref{sine-gordon-dis}.
}

{We can determine the location of the stationary points 
parametrically (in the presence of bias current but in the absence of dissipation)  based on the potential that includes the bias current $\bar{V}=V- 2 \pi \Gamma x_0$. A natural parameter in this case is $\Gamma/\varepsilon$.
Clearly, the positions of the stationary points do not change if there is a dissipation in the system.
{Figure \ref{fig_09} shows the positions of the stationary points of the potential $\bar{V}$ depending on the parameter $\Gamma/\varepsilon$, for $\varepsilon \neq 0$.
 The dashed line in the figure always marks the maximum of the potential, and hence a saddle point for the kink
dynamics, while the solid line marks the minimum of the
potential and hence the stable equilibrium of the  driven
system, participating in the pair of saddle-node
bifurcations illustrated in Figure \ref{fig_09}
and taking place at $\Gamma/\varepsilon \approx \pm 0.39$. If, for example, we consider the part of the graph for which the control parameter is positive then if $\Gamma>0$ and $\varepsilon>0$  the upper branch describes the maximum while the lower branch describes the minimum of the potential $\bar{V}$.} 

\begin{figure}[ht]
    \centering
    \subfloat{{\includegraphics[width=7.5cm]{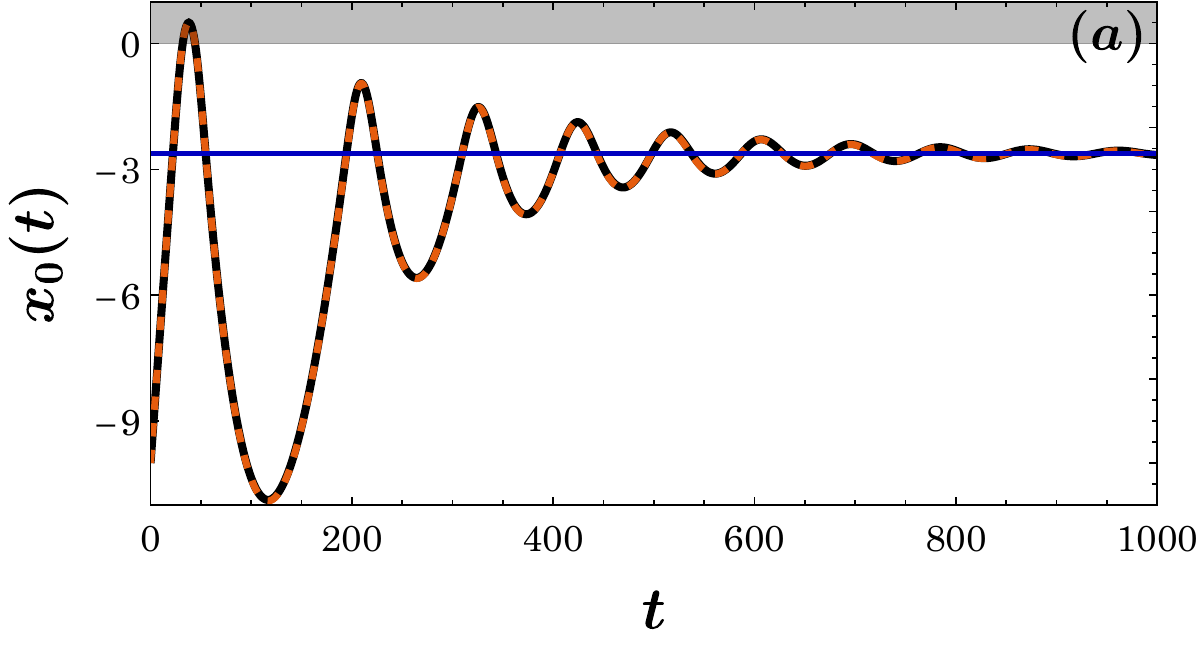}}}
    \quad
    \subfloat{{\includegraphics[width=7.65cm]{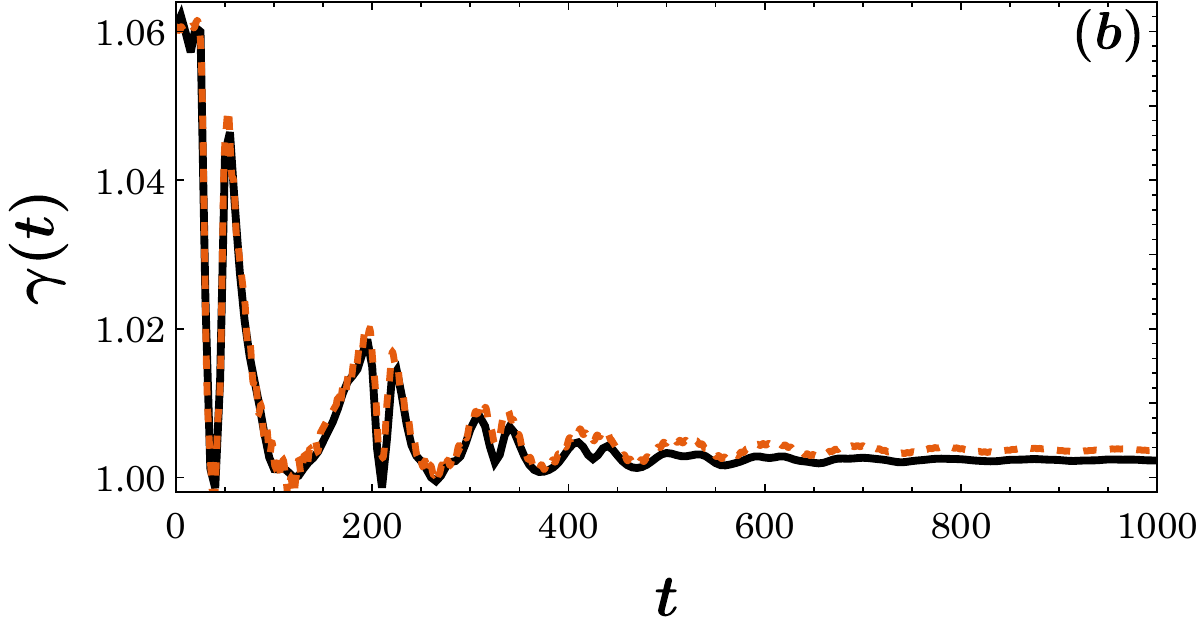}}}
    \caption{Comparison of the position of the kink {(a)} and the evolution of the $\gamma(t)$ variable {(b)} for the solution from the original field model (black line) and the model according to the equation \eqref{2dof_ansatz-dyss} (red line). The figures are prepared for $\varepsilon = 0.1$, $\alpha = 0.01$, $L=10$ and bias current $0.0045$. { The right panel {(b)} illustrates the gradual convergence of $\gamma(t)$ towards a value of $\gamma \approx 1$. For instance, at $t=1000$, $\gamma$ reaches a value of 1.002.}}
    \label{fig_10}
\end{figure}

\begin{figure}[ht]
    \centering
    \includegraphics[width=7.5cm]{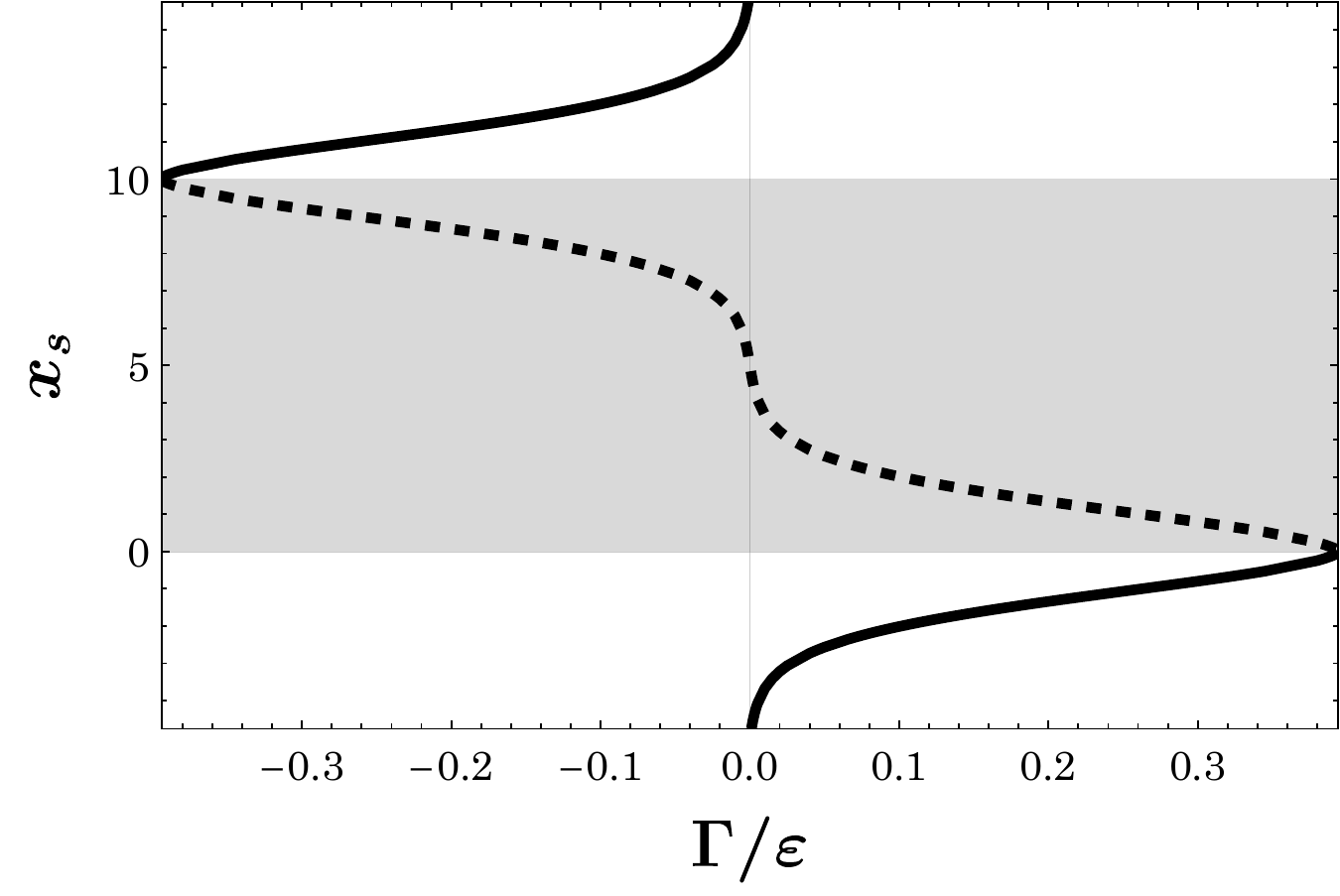}
    \caption{ Stationary points of potential $\bar{V}$. The black solid line indicates the location of the minimum, while the dashed line indicates the location of the maximum. Figure based on solutions of equation \eqref{dV} for $L=10$.}
    \label{fig_09}
\end{figure}

A simulation was also carried out for $\varepsilon=0.5$ and other parameters identical to those in Figure \ref{fig_10}. Although the agreement in the case of the $x_0$ variable was very good, when it came to the $\gamma$ variable, we noted somewhat more pronounced differences between the PDE model and the effective model at some instants of up to 15 percent. This result and further numerical analysis indicated that the main reason for the deviations is that the variation of the $g$ function is fairly fast in relation to the kink size. To verify this hypothesis, the barrier shape was relaxed. Assuming the following shape of the $g$ function
\begin{equation}
    \label{gg}
    g(x) = \tanh\left(\frac{x}{2}\right)-\tanh\left(\frac{x-L}{2}\right)
\end{equation}
the relaxation of the barrier course was achieved. At the same time, in order to ensure the same height of the barrier as before, the width of the barrier was extended, taking $L=30$. Then, $\varepsilon$ was increased to a value of $0.5$, obtaining the result presented in Figure \ref{fig_11}. As before, the agreement in the variable $x_0$ is very good {(Fig.11.a)}. On the other hand, the agreement for the variable $\gamma$ {(Fig.11.b)} has even improved compared to the result presented in Figure \ref{fig_10} (with $\varepsilon=0.1$). The other parameters that were adopted in the simulations were the dispersion coefficient $\alpha=0.01$ and the bias current $\Gamma=0.011$. These values correspond to an equilibrium (far from the defect) kink velocity of $v_s \approx 0.65$. The initial velocity of the kink was chosen according to the formula \eqref{u}. The end position of the kink was determined as before based on 
equation \eqref{dV} and corresponds to $x_0 = -4.396$. An interesting conclusion from the above simulations is that as long as the inhomogeneity is sufficiently soft then we can 
practically use the effective model to explore the model
dynamics for a wide range of system parameters.
\begin{figure}[ht]
    \centering
    \subfloat{{\includegraphics[width=7.5cm]{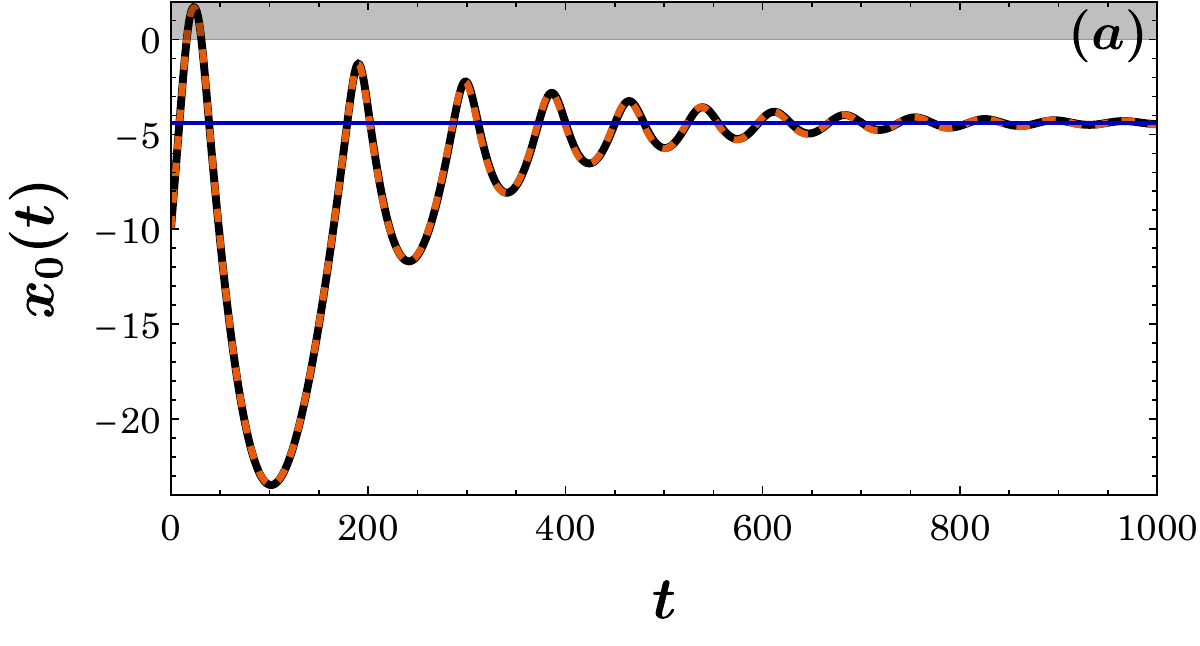}}}
    \quad
    \subfloat{{\includegraphics[width=7.65cm]{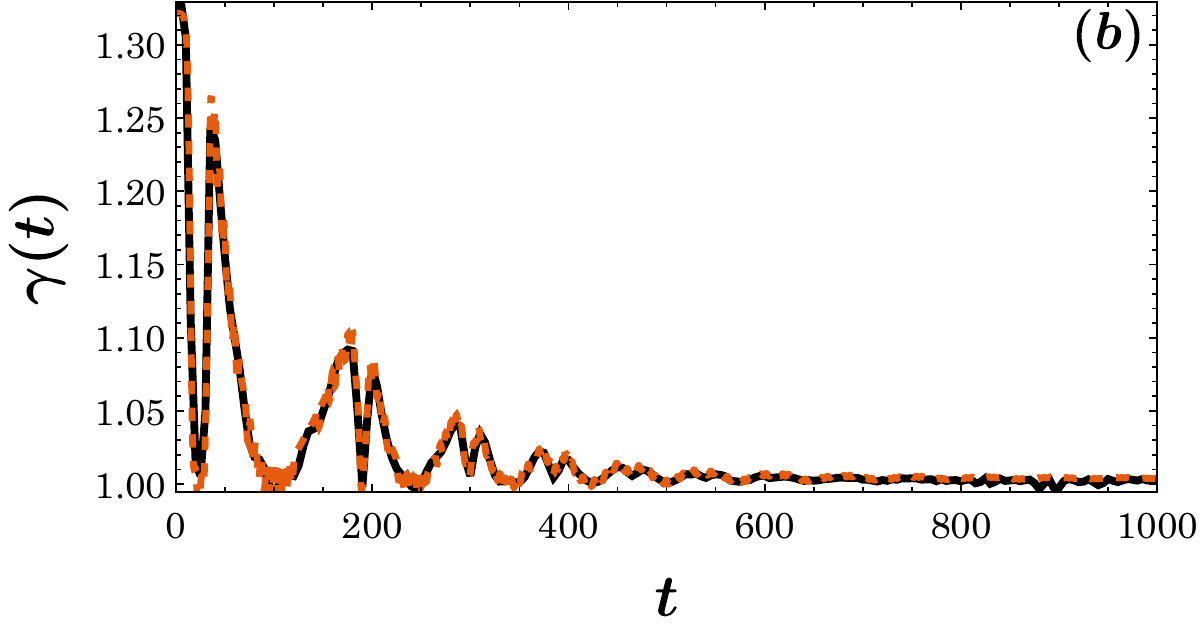}}}
    \caption{Comparison of the position of the kink {(a)} and the evolution of the $\gamma(t)$ variable {(b)} for the solution from the original PDE model (black line) and the model according to the equation \eqref{2dof_ansatz-dyss} (red line). The figures are prepared for $\varepsilon = 0.5$, $\alpha = 0.01$ and bias current  $\Gamma = 0.011$. Here $g(x)$ is given by \eqref{gg} with $L=30$. The equilibrium position is $x_0=-4.396$ (blue line). { The evolution of $\gamma(t)$ in this figure demonstrates convergence towards $\gamma \approx 1$. By $t=1000$, $\gamma$ achieves a value of 1.003.}}
    \label{fig_11}
\end{figure}

We determine the period of oscillations based on the effective model by assuming zero dissipation $(\alpha=0)$ in equations \eqref{2dof_ansatz-dyss}. To bring the equations into linearized form we must assume the values of the parameters $m$, $M$, $\kappa$ characteristic of the equilibrium position $m_s \equiv m(x_0=x_s,\gamma=\gamma_s)$,
$M_s \equiv M(x_0=x_s,\gamma=\gamma_s)$
and
$\kappa_s \equiv \kappa(x_0=x_s,\gamma=\gamma_s)$. 
Here $x_s$ and $\gamma_s$ denote equilibrium positions in the variables $x_0$ and $\gamma$ (where $\gamma_s \approx 1$). A kink placed at this position $(x_s,\gamma_s)$ remains static. The linearized equations are then as follows
\begin{equation}
\begin{gathered}
\label{eq-periods}
   M_s \delta \Ddot{x}_0 - \kappa_s \delta \Ddot{\gamma} + \partial_{x_0} V(x_s,\gamma_s) + \Omega_{x_0}^2 \delta x_0 = 2 \pi \Gamma , \\
   m_s \delta \Ddot{\gamma} - \kappa_s \delta \Ddot{x}_0  + \Omega_{\gamma}^2 \delta \gamma = 0 ,
\end{gathered}
\end{equation}
Here $\delta x_0$ and $\delta \gamma$ denote the kink deflections from the equilibrium position ie. $\delta x_0 \equiv x_0 - x_s$ and $\delta \gamma \equiv \gamma - \gamma_s$.
The other coefficients denote the second derivatives of the potential at the equilibrium position
$\Omega^2_{x_0} \equiv \partial_{x_0}^{2} V(x_s,\gamma_s)$
 and
$\Omega^2_{\gamma} \equiv \partial_{\gamma}^2 V(x_s,\gamma_s)$.
We can eliminate the bias current and the derivative of the potential in the first equation based on equation~\eqref{dV}. Subsequent to this operation, from this system of equations we can determine the approximate frequencies of oscillations
\begin{equation}\label{omega}
    \omega^2 = \frac{(M_s \Omega_{\gamma}^2 + m_s \Omega_{x_0}^2) \pm \sqrt{M_s^2 \Omega_{\gamma}^4 + m_s^2 \Omega_{x_0}^4 +2 \Omega_{\gamma}^2 \Omega_{x_0}^2(2 \kappa_s^2 - M_s m_s)}}{2 (M_s m_s - \kappa^2_s)} .
\end{equation}
The lower frequency identified as $0.071$ describes the vibration of the kink as a whole around the equilibrium position. Figure \ref{fig_12} (top left panel { 12.a}) shows the oscillation of the kink around the equilibrium position, which is located at a point with a coordinate of $-2.62$. 
This is excited by initializing the kink away from the above
equilibrium and more concretely centered around $x_0(0)=-2.72$.
This value was obtained for $\varepsilon=0.1$ and bias current $\Gamma=0.0045$. This graph was obtained based on the field equation \eqref{sine-gordon-dis} with $\alpha=0$. Due to the lack of dissipation, the oscillation of the kink position does not decrease with time. The right panel 
{ (Fig.12.b)} shows the Fourier analysis of the waveform shown in the left panel. It turns out that the obtained peak exactly corresponds to the oscillation frequency obtained based on equation~\eqref{omega}, i.e., $\omega=0.071$. 
The lower left panel of this figure { (Fig.12.c)} shows the oscillation of the variable $\gamma$. 
The bottom right panel { (Fig.12.d)} shows the result of the Fourier analysis of the waveform shown on the left. It can be seen that the value of the frequency shown in the bottom right figure ($\omega=1.109$) is very close to the higher frequency obtained from the formula~\eqref{omega}, which is $1.106$. Once again, we observe the
validation of the corresponding linearization theory
through the direct dynamical simulations.

\begin{figure}[ht]
    \centering
    \subfloat{{\includegraphics[width=7.5cm]{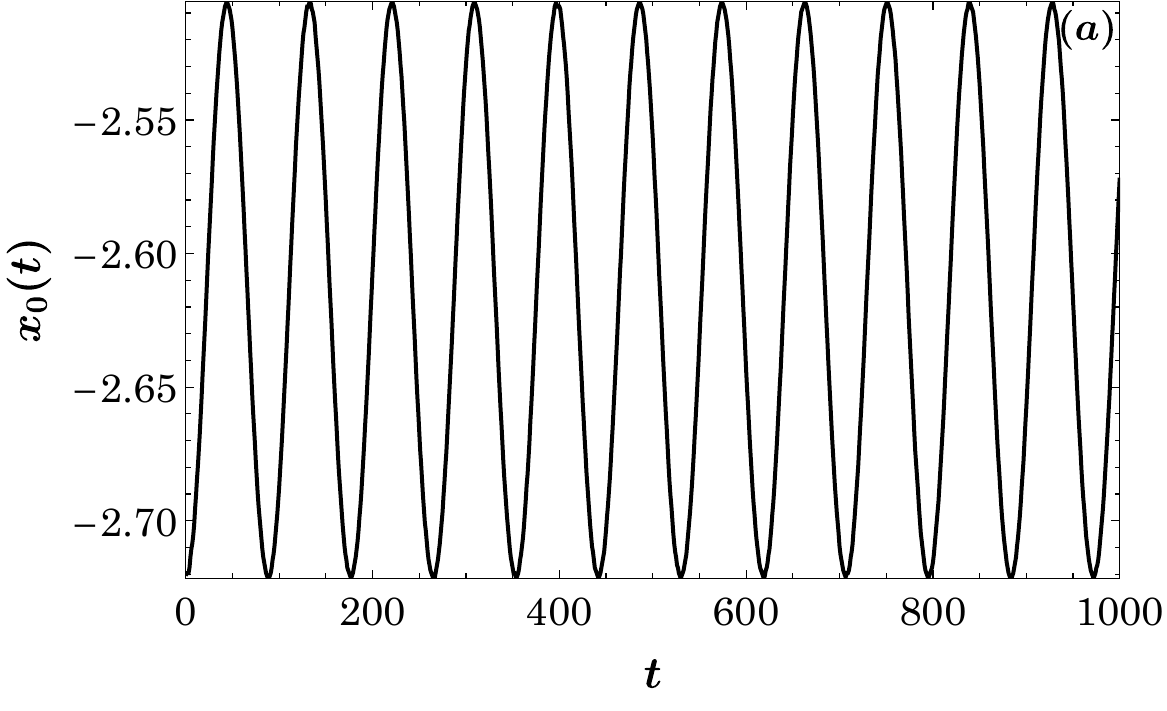}}}
    \quad
    \subfloat{{\includegraphics[width=7.15cm]{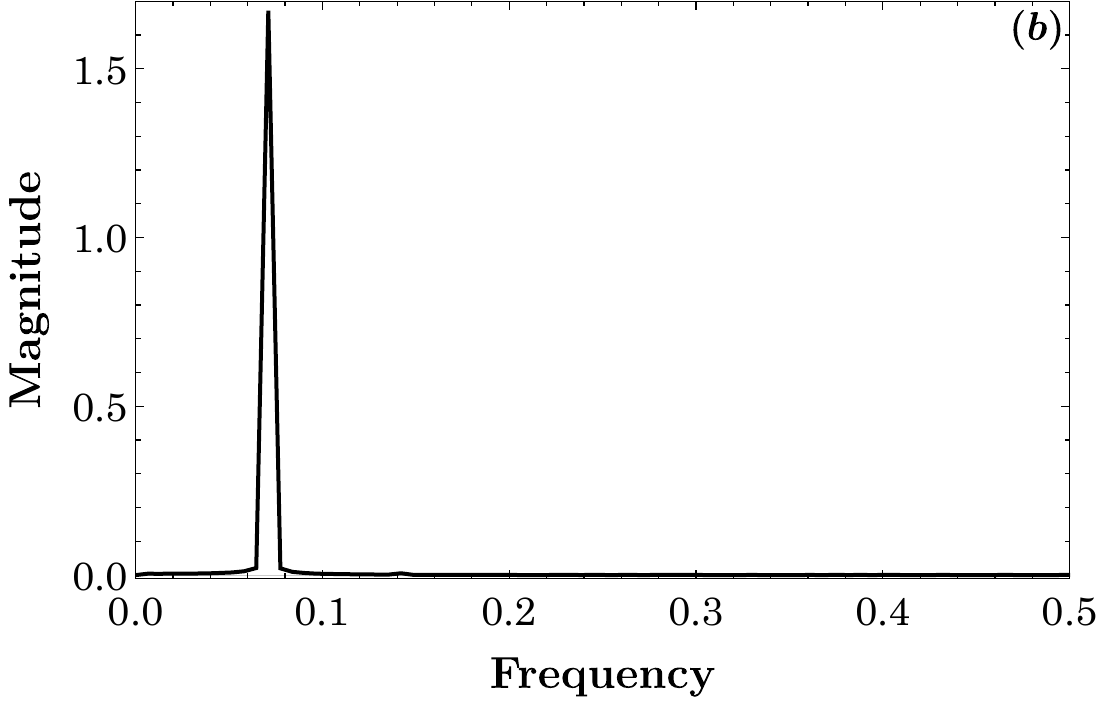}}}
    \quad
    \subfloat{{\includegraphics[width=7.5cm]{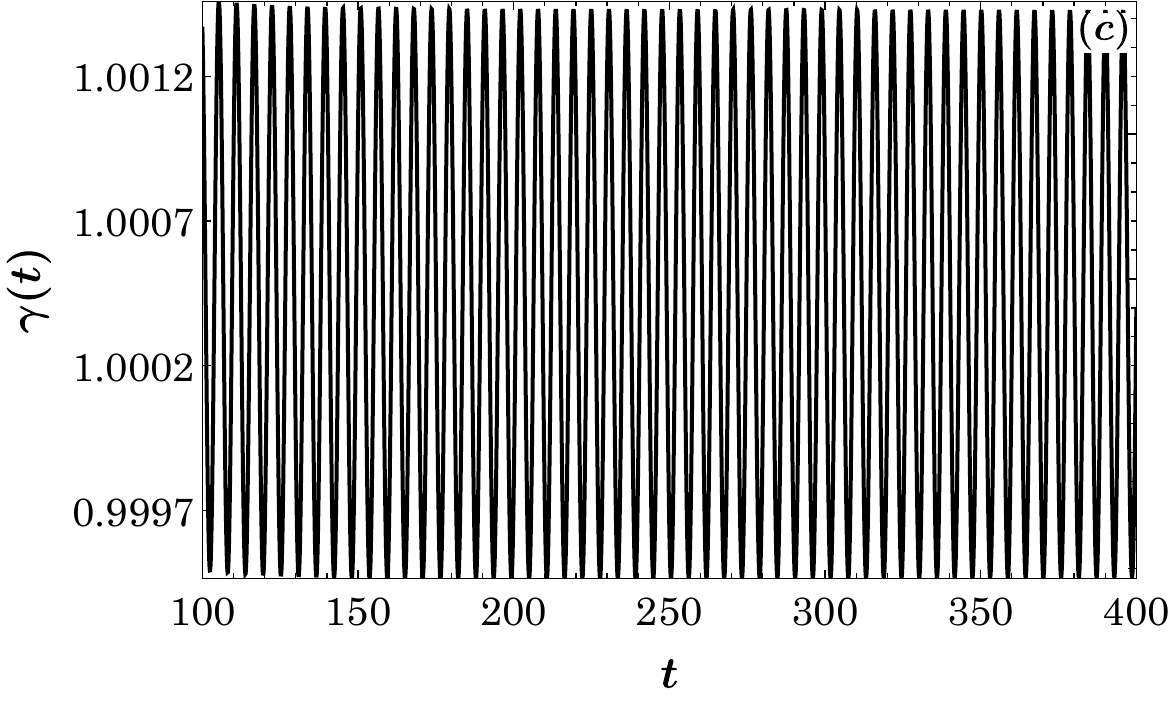}}}
    \quad
    \subfloat{{\includegraphics[width=7.15cm]{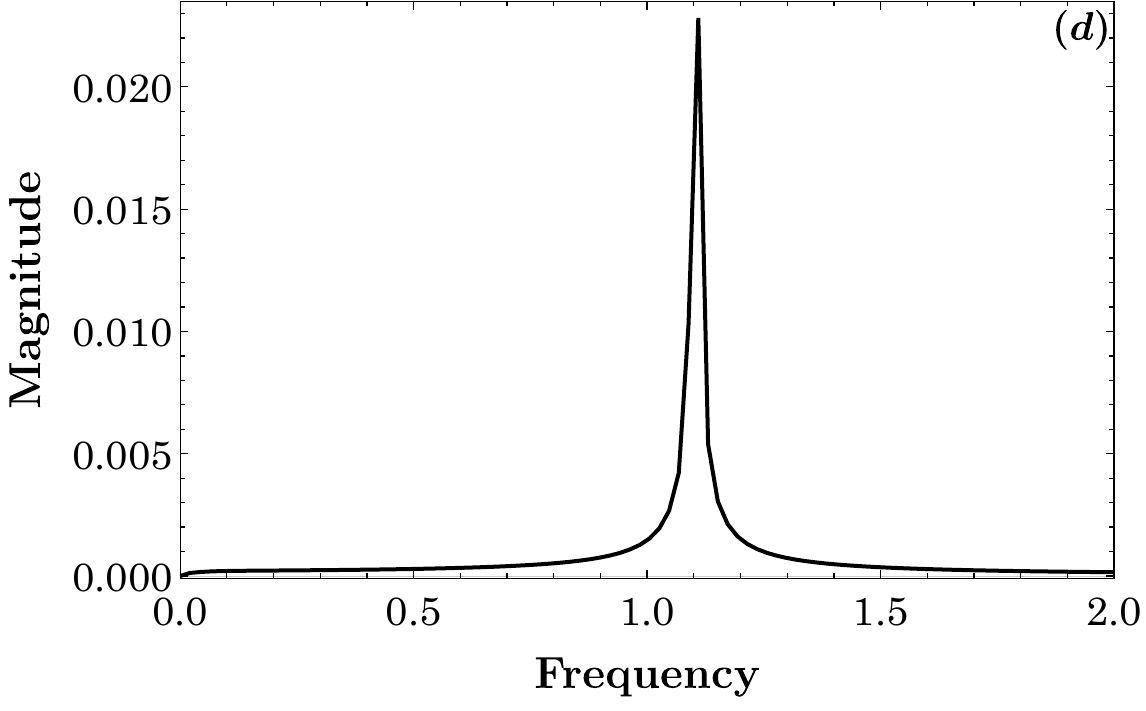}}}
    \caption{Kink oscillations around the equilibrium position obtained from the field equation \eqref{sine-gordon-dis} in the absence of dissipation ({ left panels: (a) - top, and (c) - bottom}). { Right panels (b) - top and (d) - bottom} contain the Fourier analysis of the signal presented in the left panels. 
    { Panels (a) and (b)} refer to changes in the position of the kink, while { panels (c) and (d)} refer to vibrations in the thickness of the kink, each pertaining to a 
    characteristic frequency in line with the linearization
    analysis (see text for details).
    The figures assume the parameters $\alpha=0$, $\Gamma=0.0045$ and $\varepsilon=0.1$. }
    \label{fig_12}
\end{figure}

The inclusion of dissipation in the system of effective equations \eqref{2dof_ansatz-dyss} leads to a linear approximation to a simplified system of the form
\begin{equation}
\begin{gathered}
\label{eq-periods-dis}
   M_s \delta \Ddot{x}_0 - \kappa_s \delta \Ddot{\gamma} +  \Omega_{x_0}^2 \delta x_0 = - \alpha M_s \delta \Dot{x}_0 + \alpha \kappa_s \delta \Dot{\gamma} , \\
   m_s \delta \Ddot{\gamma} - \kappa_s \delta \Ddot{x}_0  + \Omega_{\gamma}^2 \delta \gamma = - \alpha m_s \delta \Dot{\gamma} + \alpha \kappa_s \delta \Dot{x}_0 .
\end{gathered}
\end{equation}
The exponents describing the time evolution of the trajectory in the linear approximation correspond to the solution of the following quartic equation 
\begin{equation}
\begin{gathered}
\label{exponent}
 (m_s M_s - \kappa_s^2) \lambda^4 + 2 \alpha (M_s m_s 
- \kappa_s^2 ) \lambda^3 + \\ \Big( M_s
\Omega_{\gamma}^2 + m_s \Omega_{x_0}^2 + 
 \alpha^2 (M_s m_s
- \kappa^2 ) \Big) \lambda^2 + \alpha ( M_s
\Omega_{\gamma}^2 + m_s \Omega_{x_0}^2 ) \lambda +
\Omega_{\gamma}^2 \Omega_{x_0}^2 = 0
   .
\end{gathered}
\end{equation}
The real part of the exponent $\lambda$ describes the decay of the vibration over time, while the imaginary part describes its oscillation frequency. This is once again for the case
of a spiral that we have examined dynamically earlier.

If we were to consider a model with one degree of freedom then, depending on the choice of degree of freedom, we would obtain equation one (for $x_0$) or equation two (for $\gamma$) of the system of equations \eqref{2dof_ansatz-dyss} (with the redundant variable eliminated). Linearized versions of these equations are of the form \eqref{eq-periods-dis}. This means that the choice of equation and the zeroing of the redundant variable allows us to determine the exponents describing the evolution of the variable remaining.

Consider the oscillations of the thickness of the kink standing in the minimum of the potential $\delta x_0 = 0$. From the second equation of the system \eqref{eq-periods-dis} we obtain the equation for the exponent 
\begin{equation}
    m_s \lambda^2 + \alpha m_s \lambda + \Omega_{\gamma}^2 = 0 .
\end{equation}
The solutions of this equation contains the real part equal to $-\alpha/2$
\begin{equation}
\label{omega_g}
    \lambda_{\pm}^{(\gamma)} = - \frac{\alpha}{2} \pm i \frac{\sqrt{4 m_s \Omega_{\gamma}^2 - \alpha^2 m_s^2}}{2 m_s} .
\end{equation}
On the other hand, if the kink as a whole moves around the minimum without thickness oscillations $\delta \gamma = 0$, then the first equation of the system \eqref{eq-periods-dis} allows the exponent to be determined
\begin{equation}
    M_s \lambda^2 + \alpha M_s \lambda + \Omega_{x_0}^2 = 0 .
\end{equation}
The solutions to this equation are of the form
\begin{equation}
\label{omega_x}
    \lambda_{\pm}^{(x_0)} = - \frac{\alpha}{2} \pm i \frac{\sqrt{4 M_s \Omega_{x_0}^2 - \alpha^2 M_s^2}}{2 M_s} .
\end{equation}
It turns out that in the case of oscillations around the minimum, 
for suitably small dissipation $\alpha$,
both of the expressions appearing under the root in the \eqref{omega_g} and \eqref{omega_x} formulas are positive, and therefore the real parts of the exponents for both thickness and position oscillations is $\alpha/2$. This leads us to the conclusion that both types of vibration, i.e., position and thickness, are damped to the same degree.

Figure \ref{fig_13} shows the spectrum of excitations of the kink in the considered models (field model and the model with two degrees of freedom) as a function of the $\Gamma / \varepsilon$ parameter.
  On the one hand, the two top figures { (Fig.13.a and 13.b)} show the results obtained from a linearized model with two degrees of freedom \eqref{exponent} for kink position oscillations (orange points) and thickness oscillations (blue points). On the other hand, the same results are reproduced from the field model stability test (black points). It can be seen that the agreement between the two approaches is very good. Moreover, the results from equations \eqref{omega_g} and \eqref{omega_x} closely match the colored dotted lines.
 The top left panel { (Fig.13.a)} illustrates the decay of the vibration of $x_0$ and $\gamma$ in the case of the minimum of the $\bar{V}$ potential
 (represented by the quantity $-\alpha/2$ discussed above), while the top right panel { (Fig.13.b)} describes the vibration of the kink position (orange and black lines) and the vibration of its thickness (blue line). 
 It is worth noting that the top (blue) line is located in the continuous spectrum of the model in the absence of inhomogeneity, forcing and dissipation (this value is equal to one). What's more, in the model with inhomogeneity, forcing and dissipation we verified that the continuous spectrum starts at the same level, i.e., at unity, in terms of
 its imaginary part (for the sake of clarity of the figures, the continuous spectrum was not included in the figures).
 The two bottom panels  { (Fig.13.c and 13.d)} show the dependence of the real  and imaginary part of the exponent on the variable $\Gamma / \varepsilon$ in case of kink located at maximum of the potential. The figures show the sliding of the kink from the maximum, i.e., the instability of the tested configuration as a whole (orange dots in the effective model \eqref{exponent} and black dots in the field model \eqref{sine-gordon-dis} ). Moreover, the orange dots in this case also describe the result of equation \eqref{omega_x}. 
 For this equation, the expression under the square root becomes negative giving the contribution to the real part (thus becoming different from $-\alpha/2$). Moreover, the imaginary part of $\lambda$ becomes equal to zero.
 The blue dots on the left lower panel { (Fig.13.c)} represent slightly damped vibrations of the kink thickness. 
 The suppression coefficient in this case is the same as before, i.e., $ -\alpha/2$. The lower right figure { (Fig.13.d)} shows the lack of vibration of the kink slipping from the maximum (both black and blue lines are at zero). Additionally, the vibration of the thickness of the kink (blue dots) is almost identical (with an accuracy of 0.01)  to that for the minimum of the potential $\tilde{V}$. Note that the internal oscillations of the kink thickness are almost not affected by the inhomogeneities that exist in the system (the blue lines in the two lower figures are similar to those in the two upper ones).
 The corresponding kink thickness oscillations obtained from the full field model \eqref{sine-gordon-dis} confirm this interpretation (see Figure \ref{fig_14}).
 The figures were made for barrier width $L=10$ and the dissipation coefficient $\alpha=0.01$.

 { These eigenvalues were obtained through a numerical procedure involving the discretization of the steady-state problem using centered finite-differences of second-order accuracy. To ensure the reliability of the results, we employed discretizations with varying spacings, $\Delta x$, and verified the convergence of the findings. The kink profiles were computed using a Newton-Raphson iteration, which not only provided rapid convergence to the steady-state solutions but also allowed us to construct the Jacobian matrix evaluated at the converged profile. This Jacobian represents the linearization of the PDE system around the kink solution and was subsequently used to numerically compute the eigenfrequencies $\lambda$. It is important to note that in the figures, we have plotted only the discrete eigenvalues obtained through this method, ensuring that they represent the  spectrum relevant to the reduced dynamical system. These are the two pairs
    of lowest eigenvalues of the relevant PDE which have been discussed in the
    past to pertain to the translational motion of the center (lowest eigenvalue
    pair) and the internal
    mode associated with the vibration of the width of this dynamical system
    (second lowest eigenvalue pair); 
    see, e.g., \cite{Balmforth2000}.}

\begin{figure}[h!]
    \centering
    \subfloat{{\includegraphics[height=4.65cm]{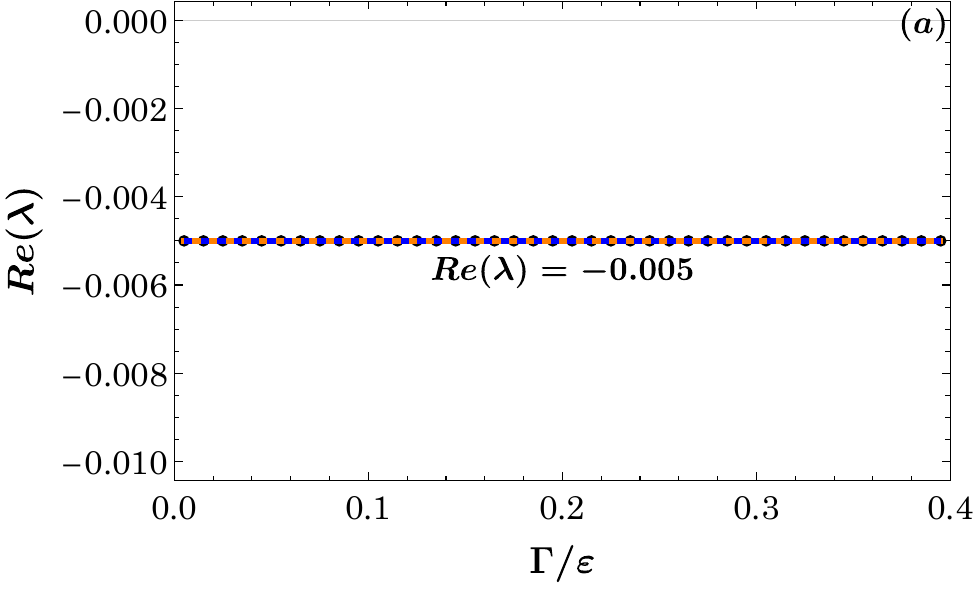}}}
    \quad
    \subfloat{{\includegraphics[height=4.5cm]{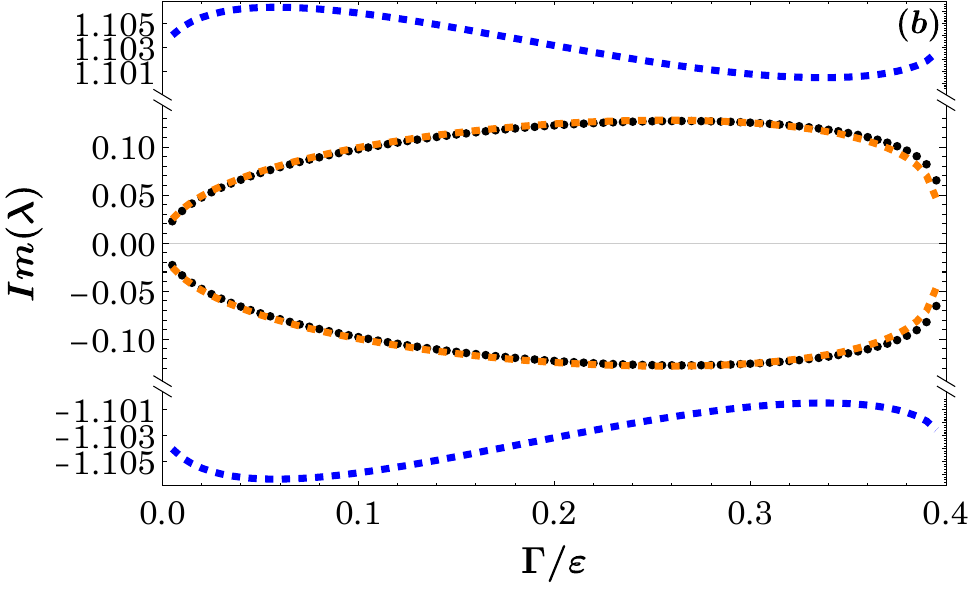}}}
    \quad
    \subfloat{{\includegraphics[height=4.5cm]{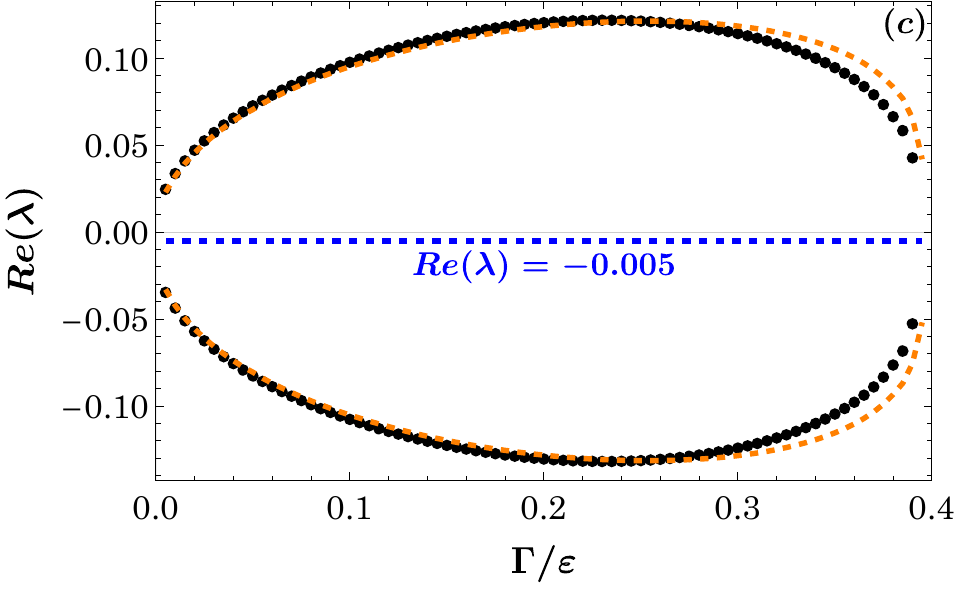}}}
    \quad
    \subfloat{{\includegraphics[height=4.5cm]{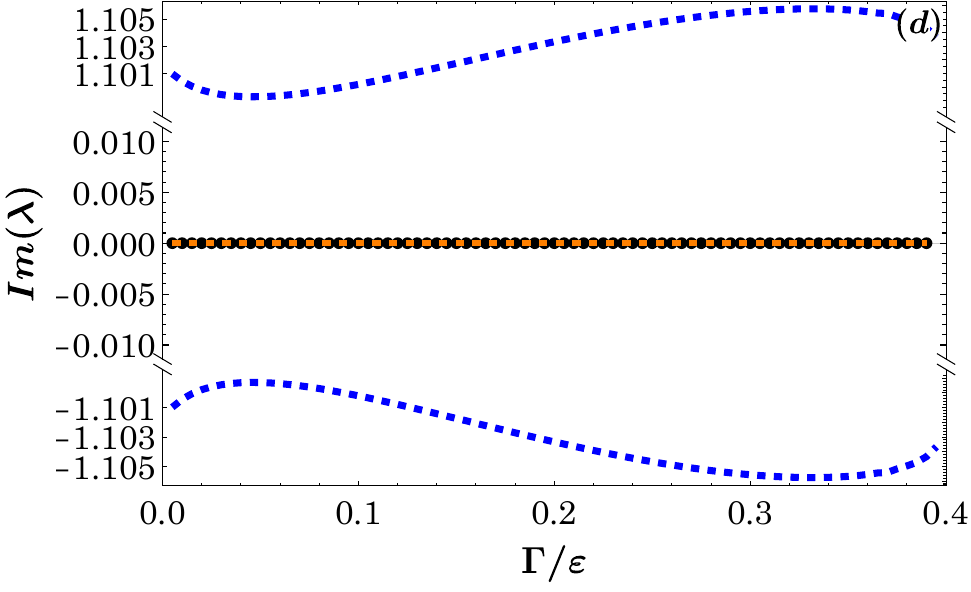}}}    
    \caption{Analysis of the stability of the kink located in the minimum and maximum of the potential $\tilde{V}$. The upper panels { (a) and (b)} show the behavior of a kink located around the minimum of the potential, while the lower panels { (c) and (d)} illustrate the instability of a kink located at the local maximum of the potential. Here we adopted the following parameters $\alpha = 0.01$ and $L=10$.}
    \label{fig_13}
\end{figure}

\begin{figure}[h!]
    \centering
    \subfloat{{\includegraphics[width=7.5cm]{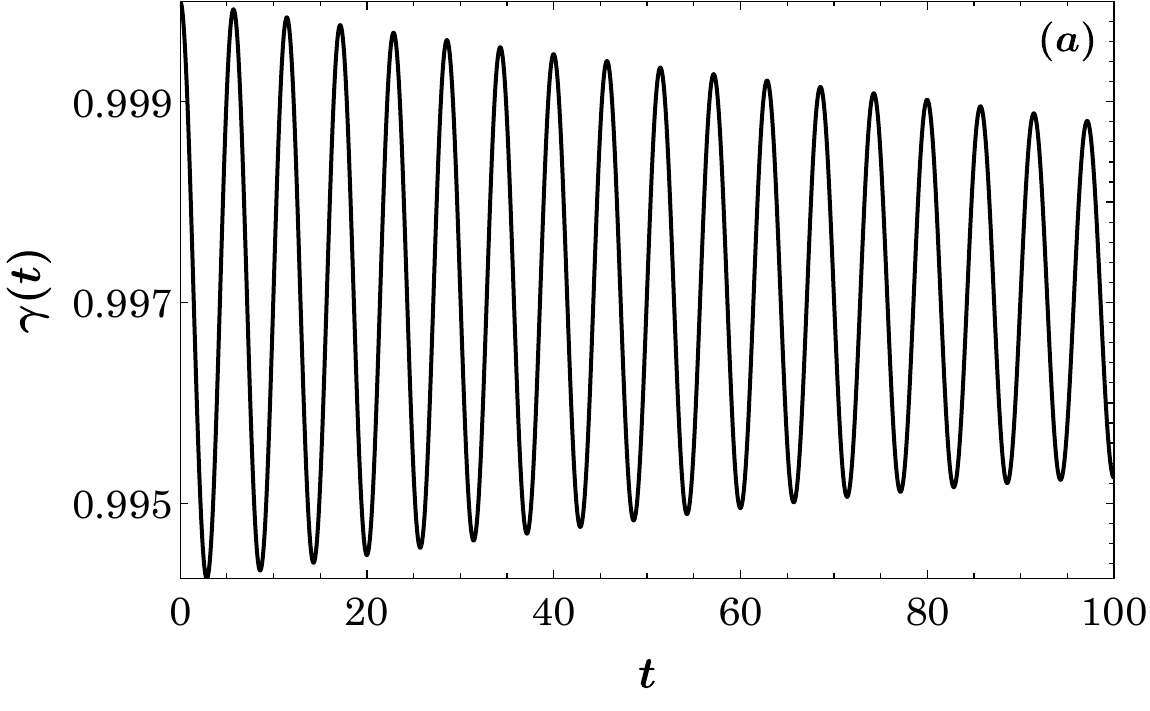}}}
    \quad
    \subfloat{{\includegraphics[width=7.15cm]{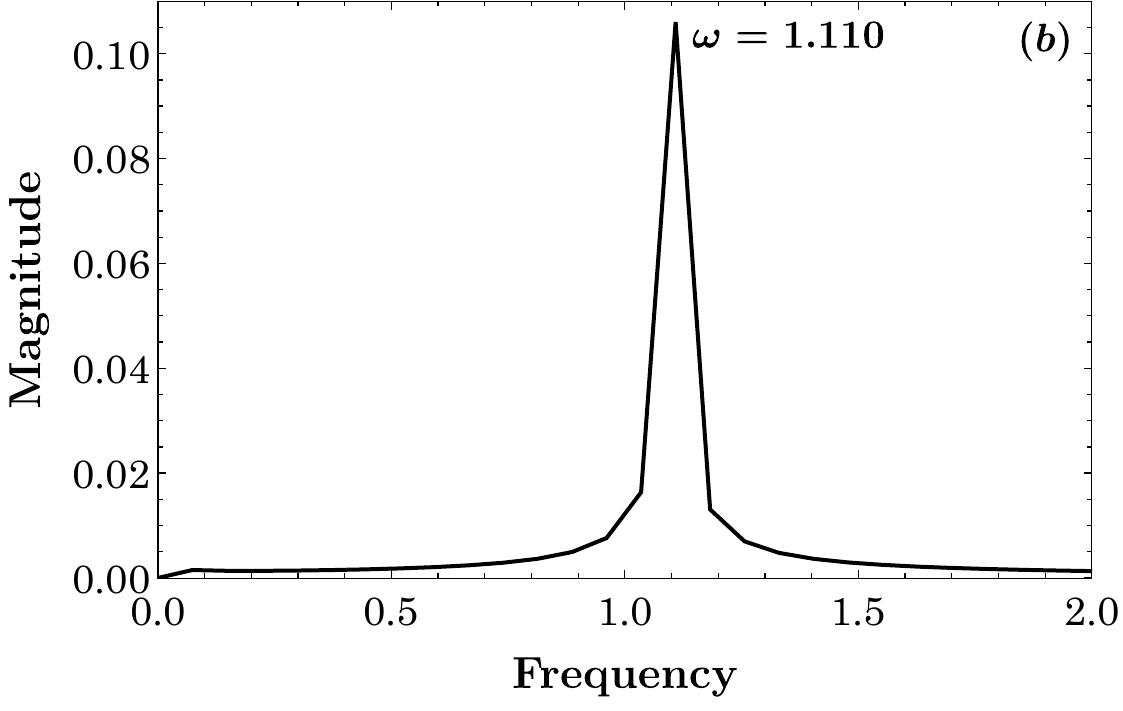}}}
    \caption{ Oscillations in the thickness of a kink that slides from a local potential maximum. The simulation was carried out for
   $\alpha=0.01$, $\Gamma=0.0045$ and $\varepsilon=0.1$ ($x_s \approx 2.62$), based on the equation \eqref{sine-gordon-dis}. }
    \label{fig_14}
\end{figure}

\FloatBarrier

\subsection{Influence of bias current on kink motion in a periodic potential}
An interesting issue not only theoretical but also one that may have technical implications is the movement of the kink in a periodic potential in the presence of dissipation and bias current. 
{Periodic perturbations of the sine-Gordon equation were studied, for example, in the context of Josephson junctions where they were associated with periodic variations in critical current, capacitance and inductance \cite{Malomed1990b, Mkrtchyan1979, Sanchez1992}.}

First, we will describe the behavior of a kink initially resting  at a node of the sinusoidal heterogeneity (at $x_0=-12$) in the presence of dissipation but in the absence of external forcing in the form of bias current. The presence of dissipation makes the kink inevitably slide toward the minimum of the sine function. In fact, we are dealing with a damped oscillation of the kink around this minimum.
The entire process is shown in Figure \ref{fig_15}. The top panel of this figure { (Fig.15.a)} shows the kink trajectory derived from the PDE model \eqref{sine-gordon-dis} (solid black line) and the effective model \eqref{2dof_ansatz-dyss}(red dashed line). The background describes the values of the $g$ function according to the legend on the right side of the figure. In the figure, the compatibility of the two models is excellent. The lower left panel { (Fig.15.b)} contains the evolution of the $\gamma$ variable. The black dotted line represents the result of the field model while the red dashed line represents the result obtained from the effective model. It can be seen that whenever the variable $x_0$ passes through the minimum of inhomogeneity, each time there is an increase in the $\gamma$ variable associated with an increase in kink velocity. 
Again, the compatibility between the two models is very good. The figure  
features the parameters selected as follows $\varepsilon = 0.1$, $x_0=-12$, $\alpha =0.021$, and $\Gamma=0$. The lower right figure { (Fig.15.c)} contains a phase portrait of the system (in cross-section $\Dot{\gamma}=0$, $\gamma=1$). In the background of this portrait is placed  the trajectory of the kink (red line), which was previously shown in the upper panel { (Fig.15.a)}.
\begin{figure}[ht]
    \centering
    \subfloat{{\includegraphics[height=4.5cm]{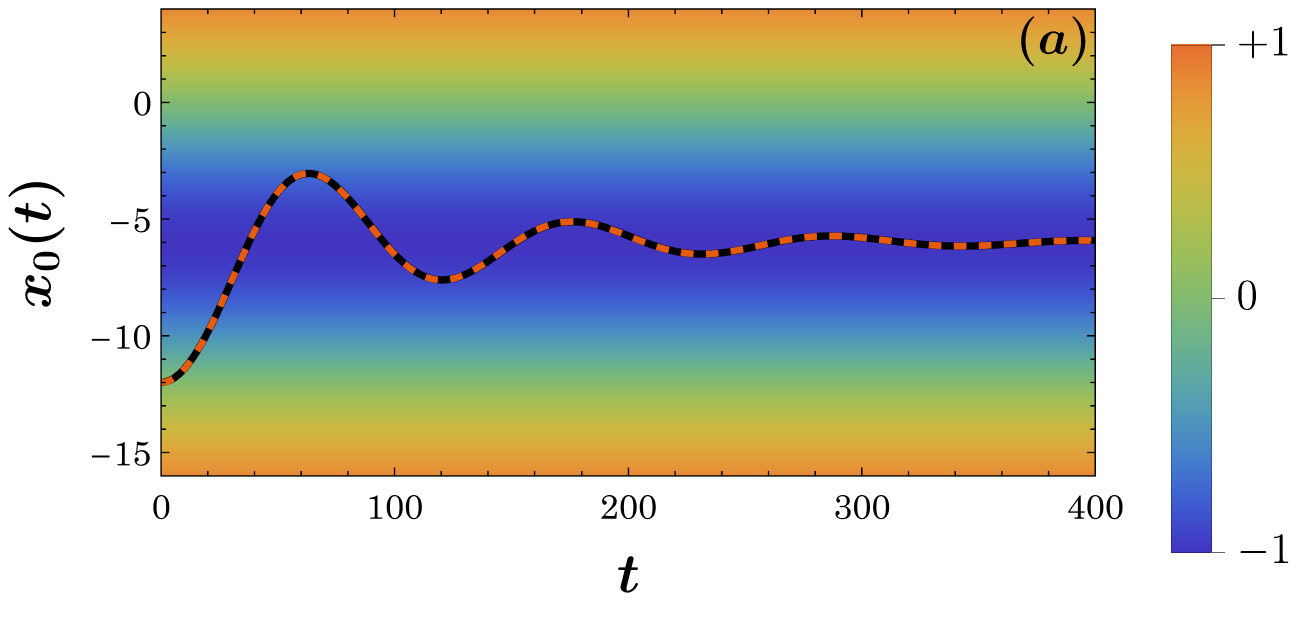}}}
    \quad
    \subfloat{{\includegraphics[height=4.5cm]{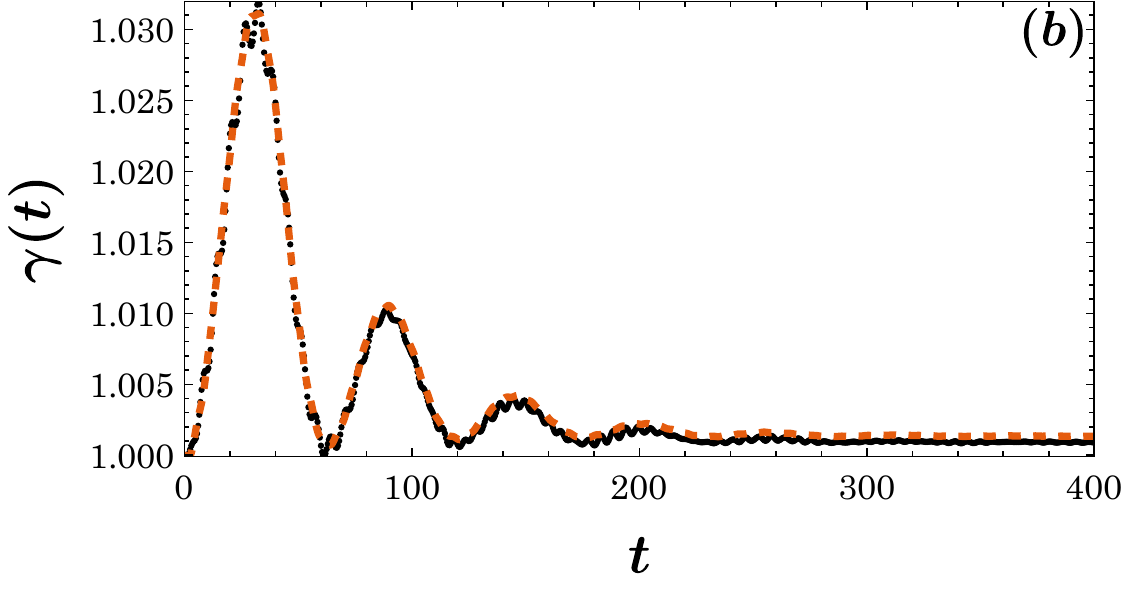}}}
    \quad
    \subfloat{{\includegraphics[height=4.5cm]{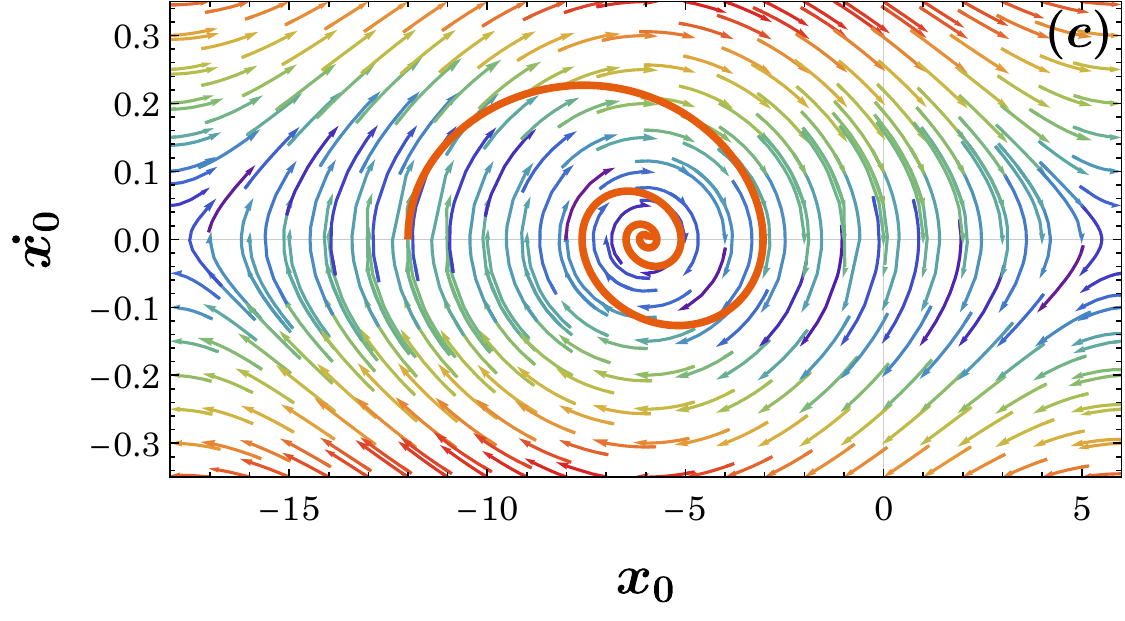}}}
    \caption{A decelerating kink eventually stopping at a potential
    minimum through the effect of dissipation. Comparison of the position of the kink { (panel (a))} and the evolution of the $\gamma(t)$ variable { (panel (b))} for the solution from the original field model (black line) and the model according to the equation \eqref{2dof_ansatz-dyss} (red line) for the periodic potential. Here $\varepsilon = 0.1$, $x_0=-12$, $\alpha =0.021$, and $\Gamma=0$. Phase diagram { (panel (c))} prepared for constant $\alpha$ and $\gamma =1$.}
    \label{fig_15}
\end{figure}

A second, even more interesting example, is the bias current-controlled  process of shifting a kink from one minimum to an adjacent minimum. This process is shown in Figure \ref{fig_16}. 
The top panel in this figure { (Fig.16.a)} shows the trajectory of the kink. As can be seen, initially the kink rests at the minimum located at $x_0=-6$. As before, the values of the g function describing the inhomogeneity are represented using the colors given in the legend on the right side of the figure. In addition, the $\varepsilon$ coefficient is equal to 0.1. One can see that, after an initial rest period, a bias current of $\Gamma=0.012$ is turned on. The time when the current takes this value in the figure is marked as a gray area. The current is turned on at $t=50$ while it is turned off at $t=200$. As a result of the external forcing, the kink is shifted to the vicinity of the adjacent minimum located at $x=18$. The bias current is then turned off. Due to the existence of dissipation in the system $\alpha=0.02$, we observe  damped oscillations of the kink around the new equilibrium position. The final state of the kink is its rest at the new minimum located at $x=18$. Note that in this figure, the black line (representing the result of the field model) is practically  identical to the red dashed line (representing the result of the effective model). 
This example shows the ability of the combination of the bias
and dissipation to engineer the motion and affect the 
eventual resting state of the kink within
a heterogeneous terrain. 
The lower left panel of this figure { (Fig.16.b)} shows the evolution of the $\gamma$ variable. As before, the gray area indicates the time when the bias current is different from zero. In the figure, the black dots indicate the result obtained from the field-theoretic PDE model \eqref{sine-gordon-dis}, while the red dots represent the result obtained from the effective ODE description \eqref{2dof_ansatz-dyss}. The agreement between the two approaches, as in previous examples,
is very good. The largest peak in this figure appears slightly after the bias current is turned off when, although the external forcing disappears, the kink slips in the direction of the inhomogeneity minimum.
In this time instance, 
the kink suffers its greatest contraction.
In the bottom right panel of this figure { (Fig.16.c)}, the trajectory from the top panel has been placed in the phase space of the described system. The cross section presented in this figure corresponds to $\gamma=1$ and $\Dot{\gamma}=0$. For the sake of convenience in interpreting this result, the part of the trajectory that corresponds to a non-zero value of the bias current is shown with a blue line. The part of the trajectory that the kink travels in the absence of external forcing is shown in red.
\begin{figure}[ht]
    \centering
    \subfloat{{\includegraphics[height=4.5cm]{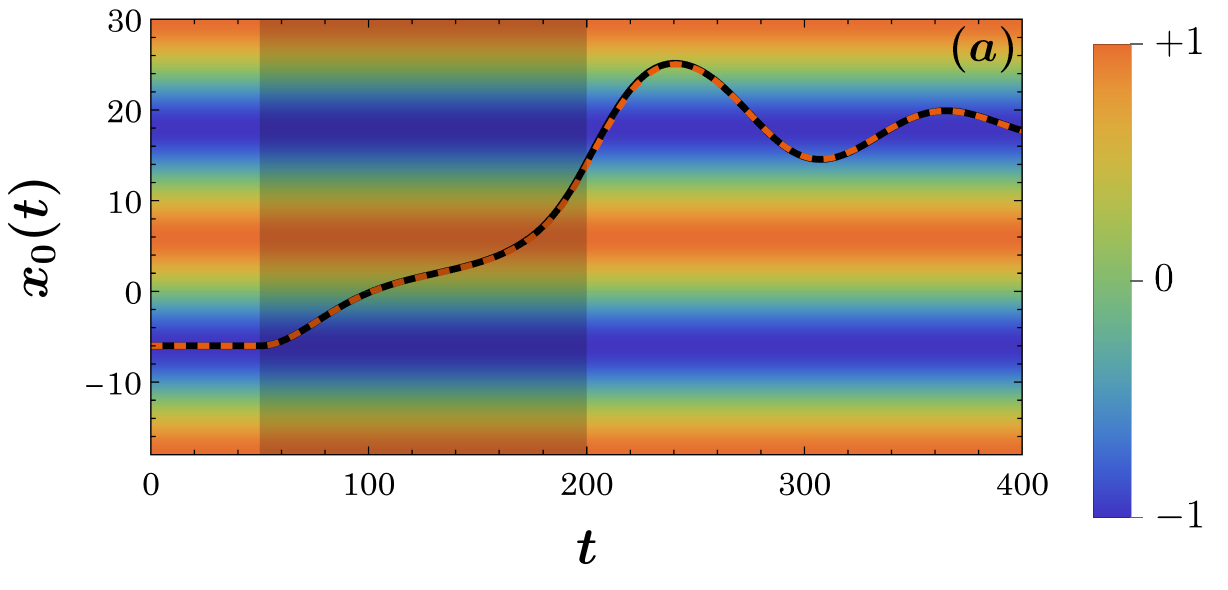}}}
    \quad
    \subfloat{{\includegraphics[height=4.5cm]{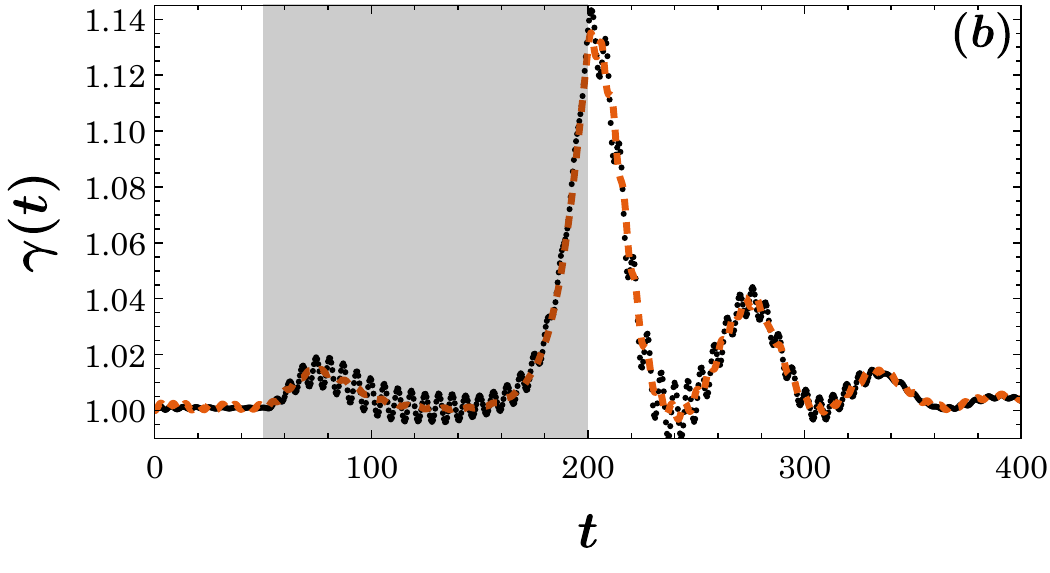}}}
    \quad
    \subfloat{{\includegraphics[height=4.5cm]{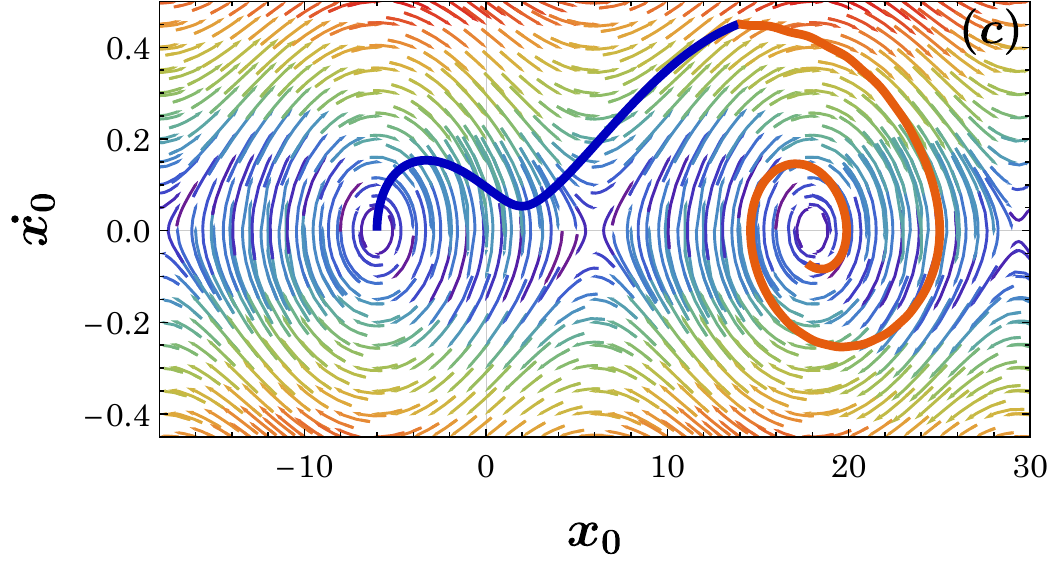}}}
    \caption{Transfer from one potential minimum to the next one. Comparison of the position of the kink { (panel (a))} and the evolution of the $\gamma(t)$ variable { (panel (b))} for the solution from the original field model (black line) and the model according to the equation \eqref{2dof_ansatz-dyss} (red line) for the periodic potential. Here $\varepsilon = 0.1$, $x_0=-6$, $\alpha =0.02$, and $\Gamma=0.012$ for $50 \leq t \leq 200$ and $\Gamma=0$ otherwise.
{
    Panel (c) shows a phase diagram prepared for constant $\gamma=1$ and $\Dot{\gamma}=0$.}
    }
    \label{fig_16}
\end{figure}

In order to check to what extent the proposed effective model \eqref{2dof_ansatz-dyss} can be considered a non-perturbative description, numerical simulations were  carried out for the parameter $\varepsilon$ describing the strength of inhomogeneity equal to $0.5$. In Figure \ref{fig_17} the bias current was also increased to $\Gamma=0.08$. The current is turned on at $t=50$ and off for $t=110$. The dispersion constant in the case under consideration was assumed to be $\alpha=0.02$.
The simulations resulted in a process that leads from a kink resting at the minimum of $x_0=-6$ to a minimum located at $x=42$. Note that these minima are not directly adjacent to each other, but are separated from each other by the minimum at $x=18$. 
As in other cases, the top panel { (Fig.17.a)} shows the trajectory of the kink against the value of the $g$ function. The trajectory determined from the PDE model (solid black line) is practically indistinguishable from the trajectory obtained from the effective model (red dashed line). In the case studied, the period in which the movement of the kink is affected by an external forcing is marked with a gray region. The time dependence of the $\gamma$ variable is shown in the lower left panel of Figure \ref{fig_17} { (Fig.17.b)}. As for the behavior of the $\gamma$ variable, it can be said that the predictions of the PDE model are consistent with those of the effective model. The only slight deviations of the effective model as far as this variable is concerned occur when there are rapid oscillations in the PDE model which are not captured in 
their detailed structure, but only in an averaged sense by the
effective ODE model. Finally, the bottom right figure { (Fig.17.c)} shows the trajectory (from the top panel) in the phase portrait of the system (in the cross section $\gamma=1$, $\Dot{\gamma}=0$). The feature that is striking in this figure is the occurrence of 
very high
velocities (they almost reach the value of 1). This means that the proposed effective description is also correct in the area of relativistic velocities.  
\begin{figure}[ht]
    \centering
    \subfloat{{\includegraphics[height=4.5cm]{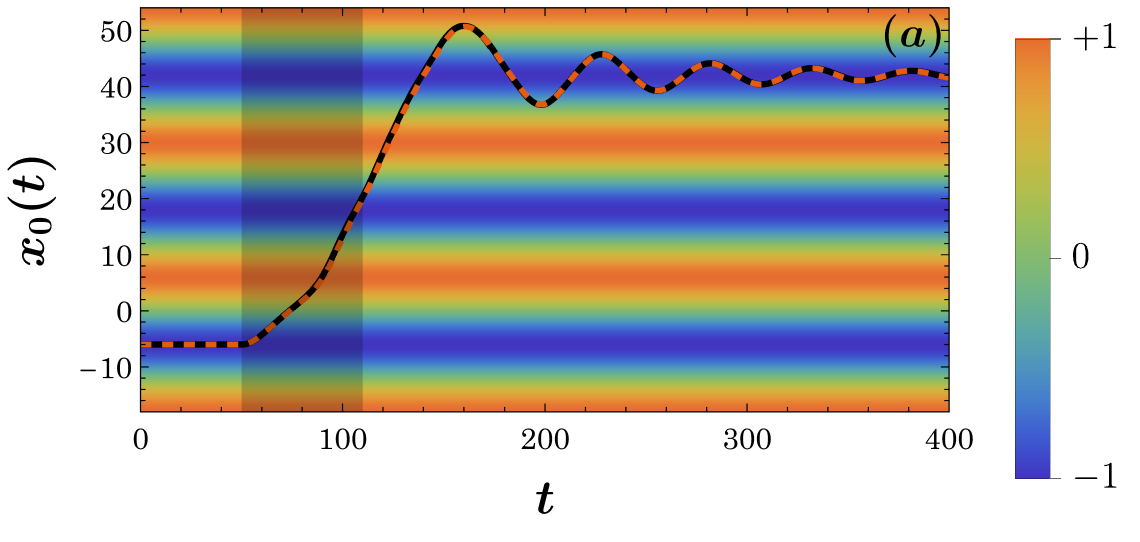}}}
    \quad
    \subfloat{{\includegraphics[height=4.5cm]{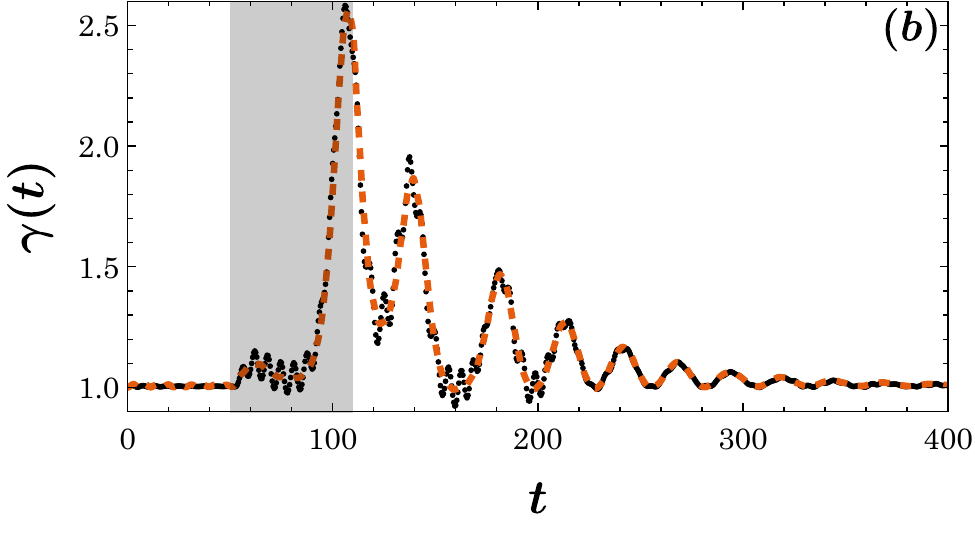}}}
    \quad
    \subfloat{{\includegraphics[height=4.5cm]{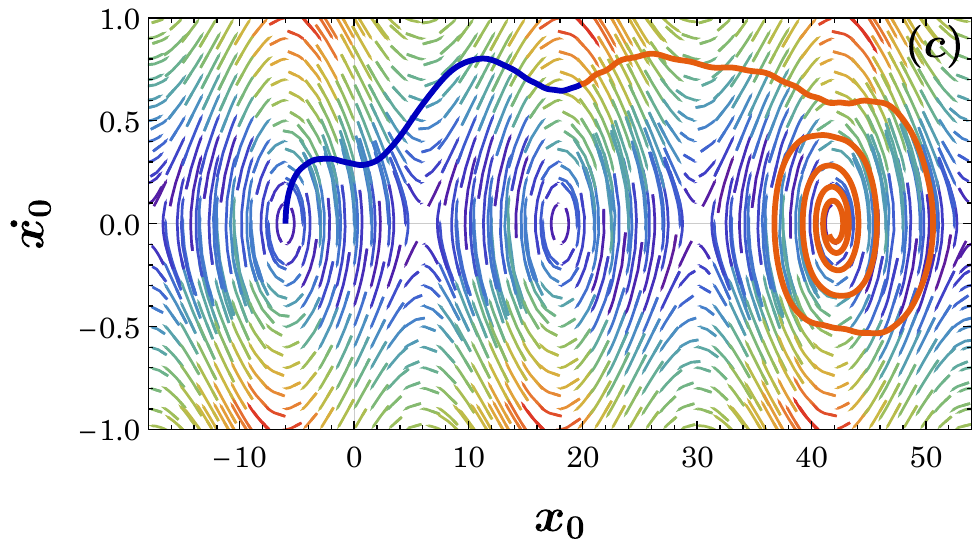}}}
    \caption{Shifting the kink from a minimum at $x_0=-6$ to a minimum that is not directly adjacent to it, i.e., $x_0=42$. { Panel (a)} shows a comparison of the position of the kink obtained from the PDE model (black line) and the effective model (red dashed line). {Panel (b)} contains an analogous comparison for the gamma variable. {Panel (c)} shows the trajectory placed on the phase portrait background, also containing the transition
    from the case with bias (blue line) to the one without (red line).
    Here $\varepsilon = 0.5$, $x_0=-6$, $\alpha =0.02$, and $\Gamma=0.08$ for $50 \leq t \leq 110$ and $\Gamma=0$ otherwise.}
    \label{fig_17}
\end{figure}

\section{Conclusions and Future Challenges}
In the present work we have revisited the problem of interaction of
a kink with a heterogeneous environment. A fundamental new insight
of the present work concerns a modified ansatz (accounting for
the heterogeneity and reducing the side effects of using the
original, unperturbed sine-Gordon kink) and the inclusion of
a variable describing the inverse width of the kink. This has
enabled the characterization of both the { conservative} case, but
also the damped driven one at a degree of accuracy that is substantially
higher than those of earlier works. Indeed, we have been able to 
realize not only the case of a localized heterogeneity but also
that of a periodic heterogeneity as a prime example towards
describing arbitrary functions, given their decomposition into
Fourier modes. We have utilized as diagnostics both the position
of the center, as well as the evolution of the inverse
width and we have offered convincing evidence that {\it both} of
them are captured far more accurately than ever before. Indeed,
the description is so efficient, that it seems plausible 
that one could use the effective ODE model {\it instead of} the PDE
for wide parametric ranges. An additional possibility that the
more complex (periodic in our case) landscape provided was that
of potentially manipulating the kink at will and controlling its
trajectory, and especially its ultimate fate/rest position, depending
on the turning on/off of the external bias drive.

Naturally, this work paves the way for further future considerations,
especially given the level of control of kink dynamics that is 
analytically and numerically possible. A natural extension concerns
the dynamics of the same model around two-dimensional impurities
in both the radial and in Cartesian directions in the spirit of
the work~\cite{Gatlik2024}. Yet another question concerns the
potential generalization of the present considerations towards a 
palette of additional, non-integrable Klein-Gordon models to explore
the potential additional dynamics therein. An example is the case
of the $\phi^4$ model where additional internal modes of the kinks 
exist~\cite{p4book}. Finally, other directions that may be interesting
concern the interaction of a kink (e.g., in the form
of a dislocation) with a random field of stationary defects
discussed, e.g., in the work of~\cite{PLC} and motivated
by intriguing physical applications, including the discussion therein of the
famous Portevin-Le Chatelier effect. Such studies are currently in progress
and will be reported in future publications.

{
On the other hand, a more precise description of the considered system could be obtained by including the surface impedance of the junction. Taking this effect into account essentially means,
at a modeling level, 
extending the field equation with a term of the form $ -  \sigma \partial_t \partial^{\,2}_x \phi$, where $\sigma$ is a small parameter. The well-posedness of such problem on a finite interval, for various boundary conditions, have been proved in the article \cite{Angelis2013}. The description of such a generalized setting is another direction for further study.}

\section*{Acknowledgments}
This material is based upon work supported by the U.S. National Science Foundation under the awards PHY-2110030 and DMS-2204702 (PGK). The work at AGH University was supported by the National Science Centre, Poland, Grant OPUS: 2021/41/B/ST3/03454, and the “Excellence Initiative-Research University” program for AGH University of Krakow (JG).

\section{Appendix A}
After integration, the parameters appearing in the equations of motion can be written in the form of
\begin{equation}
\begin{gathered}
\label{integrals-fin}
    M = \frac{\pi^2}{6 \gamma} \sqrt{{\cal F}(x_0)} \left( \frac{\partial_{x_0}{\cal F}(x_0)}{{\cal F}(x_0)}\right)^2 + \frac{8 \gamma}{\sqrt{{\cal F}(x_0)}},\\
    m = \frac{2 \pi^2}{3 \gamma^3} \sqrt{{\cal F}(x_0)},\\
    \kappa = \frac{\pi^2}{3 \gamma^2} \frac{1}{\sqrt{{\cal F}(x_0)}} \, \partial_{x_0} {\cal F}(x_0),\\
    V = \frac{4 \sqrt{{\cal F}(x_0)}}{\gamma} +\frac{2 \gamma^2}{{\cal F}(x_0)} \int_{-\infty}^{+\infty} dx \sech^2 \xi \, {\cal F}(x) .
\end{gathered}
\end{equation}
As can be seen in the case of potential, its explicit analytical form is determined by the choice of the analytical form of the ${\cal F}(x)$ function.
{\section{Appendix B}
In this appendix, we compare on the ground of the field model \eqref{sine-gordon} the positions of the point $\phi=\pi$ and the center of mass of the kink. We considered the interaction of a kink with a potential barrier. We carried out a comparison for both the case when there is a reflection of the kink from the barrier (Fig. 18.a) and the case when the kink passes over the barrier (Fig. 18.b).
The gray area in the figures represents the position of the barrier. The black line represents the position of the kink's center of mass, while the red dashed line describes the location of the kink defined by the equation $\phi=\pi$. It can be seen that in both drawings the two lines overlap. In the first case, the kink's initial velocities are quite low (Figure \ref{fig_18} is for illustrative purposes only, simulations were performed for various speeds), while in the second case they were higher. Moreover, increasing the velocity does not lead, when passing over the barrier, to a difference in the results obtained. Interestingly, slight differences appeared for speeds in the initial speed range, which includes the critical velocity. In the case of a barrier with a height of $\varepsilon=0.1$, these speeds were in the range of $0.39$ to $0.44$. }

\begin{figure}[h!]
    \centering
    \subfloat{{\includegraphics[width=7.5cm]{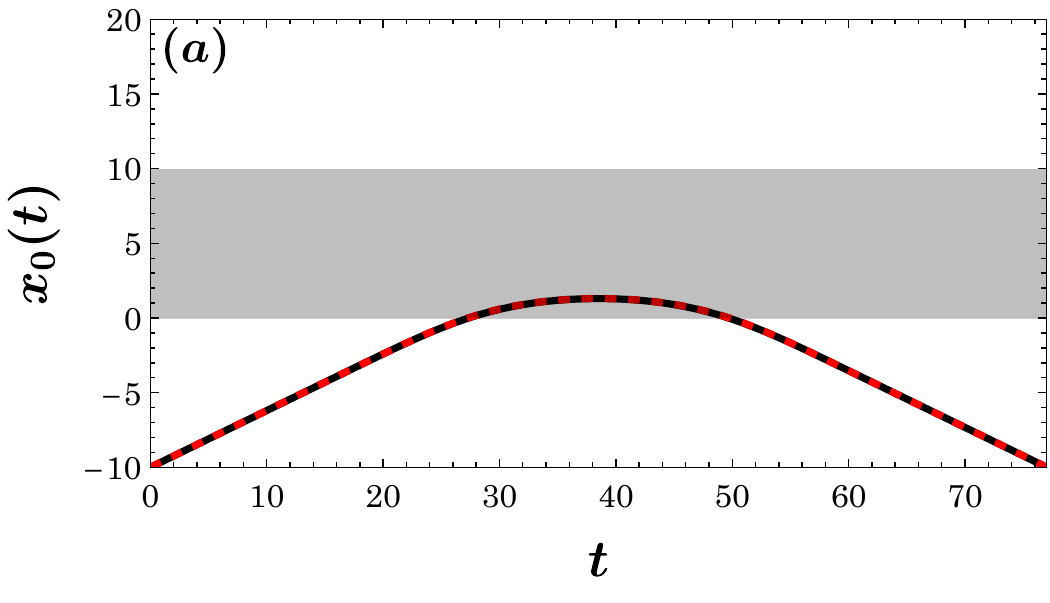}}}
    \quad
    \subfloat{{\includegraphics[width=7.15cm]{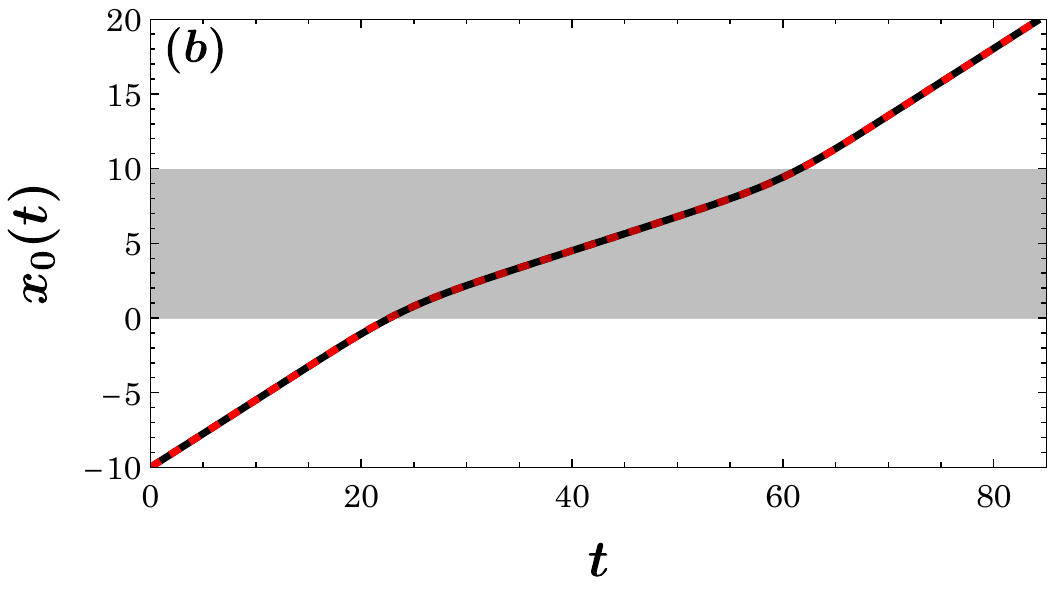}}}
    \caption{(a) Reflection of the kink from the barrier at initial velocity $v_0=0.38$. (b) Passage of the kink over the barrier at velocity $v_0=0.45$. In both cases, the height of the barrier is $\varepsilon=0.1$ and its width is $L=10$.
    }
    \label{fig_18}
\end{figure}

\FloatBarrier
\printbibliography

\end{document}